\documentclass[12pt]{emulateapj-rtx4}
\usepackage{txfonts}
\renewcommand{\cite}{\citealp}

\begin{document}

\title{On a new theoretical framework for RR Lyrae stars I: the metallicity dependence}

 \author{M. Marconi\altaffilmark{1}, 
G. Coppola\altaffilmark{1}, 
G. Bono\altaffilmark{2,3}, 
V. Braga\altaffilmark{2,3},
A. Pietrinferni\altaffilmark{4}, 
R. Buonanno\altaffilmark{2,4}, 
M. Castellani\altaffilmark{3}, 
I. Musella\altaffilmark{1},
V. Ripepi\altaffilmark{1},
R. F. Stellingwerf\altaffilmark{5}
}   

\altaffiltext{1}{INAF-Osservatorio astronomico di Capodimonte, Via
  Moiariello 16, 80131 Napoli, Italy; marcella.marconi@oacn.inaf.it}
\altaffiltext{2}{Dipartimento di Fisica - Universit\`a di Roma Tor Vergata, Via della Ricerca Scientifica 1; giuseppe.bono@roma2.infn.it}
\altaffiltext{3}{INAF-Osservatorio Astronomico di Roma, Via Frascati 33, 00040 Monte Porzio Catone, Italy; marco.castellani@oa-roma.inaf.it}
\altaffiltext{4}{INAF-Osservatorio Astronomico di Collurania, Via M. Maggini, Teramo, Italy; adriano@oa-teramo.inaf.it; roberto.buonanno@oa-teramo.inaf.it}
\altaffiltext{5}{Stellingwerf Consulting, 11033 Mathis Mtn Rd SE, 35803 Huntsville, AL USA; rfs@swcp.com}

 \begin{abstract}

We present new nonlinear, time-dependent convective hydrodynamical 
models of RR Lyrae stars computed assuming a constant helium-to-metal 
enrichment ratio and a broad range in metal abundances ($Z=0.0001$--$0.02$).  
The stellar masses and luminosities adopted to construct the pulsation 
models were fixed according to detailed central He burning Horizontal Branch evolutionary
models. The pulsation models cover a broad range in stellar luminosity and 
effective temperatures and the modal stability is investigated for both 
fundamental and first overtones. We predict the topology 
of the instability strip as a function of the metal content and new analytical relations for the edges of the instability strip in 
the observational plane. Moreover,  a new analytical relation
to constrain the pulsation mass of double pulsators as a function of the 
period ratio and the metal content is provided.   
We derive new Period-Radius-Metallicity relations for fundamental and first-overtone pulsators. 
They agree quite well with similar empirical and theoretical relations in the literature.  
From the predicted bolometric light curves, transformed into optical 
($UBVRI$) and near-infrared ($JHK$) bands, we compute the intensity-averaged 
mean magnitudes along the entire pulsation cycle and, in turn, new and homogenous metal-dependent ($RIJHK$) Period-Luminosity 
relations. Moreover, we compute new dual and triple band optical, 
optical--NIR and NIR Period-Wesenheit-Metallicity relations. 
Interestingly, we find that the optical Period-W($V,B-V$) is independent 
of the metal content and
that the accuracy of individual distances is a balance between the  
adopted diagnostics and the precision of photometric and 
spectroscopic datasets.  
 \end{abstract}


\section{Introduction} 

RR Lyrae stars (RRLs) are low--mass helium burning stars playing a crucial role 
both as standard candles and tracers of old (t$>$10 Gyr) stellar 
populations~\citep{b11,marconi12}. The RRLs have been detected in different 
Galactic~\citep[see e.g.][and references therein]{VZ06,Z14,drake13,petru14} and 
extragalactic~\citep[see e.g.][and references therein]{moretti09,sos09,fio10,sosz10,fio12,cusano13} 
environments, including a significant fraction 
of globular clusters~\citep[][]{coppola11, dicrisci11,k13,ku13}.
One of the key advantage in using RRLs is that they can be easily identified 
thanks to the shape of their light curves, the luminosity amplitudes and 
pulsation periods. They are also relatively bright, typically 3.0/3.5 mag 
brighter than the Main Sequence Turn Off (MSTO) 
stars~\citep[see e.g.][]{piersimoni02,ripepi07,coppola11,coppola13,braga14}.  
The RRLs have been popular primary distance indicators thanks to the 
relation between the absolute visual magnitude and the iron abundance 
$[Fe/H]$~\citep[][]{caputo00a,caccclem03,b11,marconi12}.
The intrinsic errors and systematics affecting distances based on 
this relation have been widely discussed in the 
literature~\citep[][]{cassisi98,caputo00a,dmc04,cass08,marconi09,marconi12}
However, RRLs have been empirically recognised to obey to a 
Period-Luminosity (PL) relation in the near-infrared (NIR) 
bands~\citep{longmore86,longmore90}. The physical bases, the key advantages
in using NIR PL relations to estimate individual distances together 
with their metallicity dependence have been extensively discussed 
in the literature both from the observational and the theoretical 
point of view~\citep[][]
{bono01,bono02a,b03,catelan04,dallora04,s06,marconi09,marconi12,coppola11,b11,braga14}. 
Here, we only mention that the NIR PL relations are marginally 
affected by uncertainties in reddening correction and by evolutionary 
effects~\citep[][]{b03}. In spite of these unquestionable advantages   
the NIR PL relations might also be prone to systematic errors.  
a) The reddening correction becomes a thorny problem if the targets 
are affected by differential reddening. This is the typical problem 
in dealing with RRL distances in the low-reddening regions of the Galactic 
Bulge~\citep{mats13} and in the inner Bulge~\citep{sosz14}. 
b) Even if the width in magnitude of the RRL instability strip at fixed period
is almost halved when moving from the I-band to the K-band, the PL relations are intrinsically a statistical diagnostics to estimate the distances, since the 
width in temperature is neglected. This means that the intrinsic dispersion of 
the PL relations, even in the NIR bands, is still affected by the width in 
temperature of the instability strip.  

To overcome the above problems it has been empirically suggested to use the 
optical and the optical-NIR Period-Wesenheit (PW) 
relations~\citep[][Coppola et al. 2015, in preparation]{dmc04,braga14}. 
The recent literature concerning the use of the reddening free Wesenheit magnitudes 
is quite extensive, but it is mainly focussed on classical 
Cepheids~\citep[][]{ripepi12,riess12,fiorentino13,inno13}.
The two main advantages in using the PW relations are: 
a) they are reddening free by construction; 
b) they mimic a Period-Luminosity-Color (PLC) relation. 
This means that individual RRL distances 
can be estimated with high accuracy, since they account for the position 
of the object inside the instability strip. Therefore, the PW relations 
have several advantages when compared with classical distance diagnostics. 
However, their use to estimate distances of field and cluster RRLs has been 
quite limited. They have been adopted by~\citet{dmc04,braga14} to estimate 
the distance of a number of Galactic GCs and by~\citet{sosz14,petru14}
to derive the distances of RRLs in the Galactic Bulge.
The inferred individual and average distances, based on the two quoted
methods (PL and PW relations), can be used to derive the 3D structure 
of the investigated stellar systems, as well as to trace radial trends 
across the Halo and tidal stellar streams~\citep[e.g.][]{sosz10,cusano13,moretti14,petru14,sosz14}.

The lack of detailed investigations concerning pros and cons of RRL PW relations 
also applies to theory. Indeed, we still lack detailed constraints on their 
metallicity dependence and on their intrinsic dispersion when moving from optical, 
to optical-NIR  and to NIR PW relations.   

The above limitations became even more compelling, during the last few years, 
thanks to several ongoing large-scale, long-term sky variability surveys. 
The OGLE~IV collaboration already released $VI$-band photometry for more
than 38,000 RRLs in the 
Galactic Bulge~\citep{petru14,sosz14}. They plan to release similar 
data for RRLs in the Magellanic Clouds (MCs). The ASAS collaboration is still 
collecting $VI$-band photometry for the entire southern sky~\citep{po14}.      
Multi-band photometry for a huge number ($\sim$23000) of southern field RRLs has 
also been released by CATALINA~\citep{drake13,tc14}. Similar 
findings have been provided in the NIR bands by VVV for RRLs in the
Galactic bulge~\citep{minniti14}  and by VMC~(Moretti et al. 2015, 
in preparation) for RRL in the MCs. New optical catalogs have also been released 
by large optical surveys such as SDSS and Pan-STARRS1~\citep{abbas14} and 
UV surveys such as GALEX~\citep{gezari13,kinman14}. New detections of field 
RRLs have also been provided, as ancillary results, by photometric surveys 
interested in the identification either of moving 
objects~\citep[LINEAR,][]{sesar11,sesar13} or of near earth 
objects~\citep{miceli08} or of transient phenomena (ROTSE, Kinemuchi
et al. 2006; PTF, Sesar et al. 2014) or of the identification of stellar streams in 
the Galactic Halo~\citep[][]{VZ06,Z14}  

The above evidence indicates that a comprehensive theoretical investigation 
addressing PL and PW relation in optical and NIR bands is required.  
Our group, during the last twenty years, constructed a detailed evolutionary 
and pulsation scenario for RRLs~\citep[see e.g.][]{bs94,bccm95a,bccm97c,m03,dmc04,m11}. We have computed several grids of Horizontal 
Branch (HB) models and pulsation models 
accounting for a wide range of chemical compositions (iron, helium), stellar masses 
and luminosity levels. The above theoretical framework was adopted to compare 
predicted and observed properties of RRLs in a wide range of stellar environments: 
globular clusters (GCs)~\citep[$\omega$ Cen,][]{m11}, nearby dwarf
galaxies~\citep[e.g. Carina and Hercules;][]{coppola13,stetson14,musella12},
the Galactic bulge~\citep{b97b,gub08}   
and the Galactic halo~\citep{fiorentino14}.    

However, the quoted theoretical framework was built using a broad range of 
evolutionary prescriptions concerning stellar masses, luminosity
levels and their dependence 
on chemical composition~\citep[][]{cassisi98,cassisi04,piet04,piet06}.
Therefore, we decided to provide a new spin on the evolutionary and pulsation 
properties of RRLs covering simultaneously a broad range in chemical compositions, 
stellar masses and luminosity levels. The current approach, when compared with the 
quoted pulsation investigations, has several differences.     
We take account of seven $\alpha$-enhanced chemical compositions ranging 
from very metal-poor ($Z=0.0001$) to the standard solar value ($Z=0.02$). 
The adopted stellar masses and luminosity levels, at fixed chemical 
composition, are only based on evolutionary prescriptions, rather than
relying on a grid of masses and luminosities encompassing evolutionary values.  

The structure of the paper is the following. In \S2 we present the new 
theoretical framework and discuss the input physics adopted to construct 
both evolutionary and pulsational models. \S3 deals with the new pulsation 
relations for fundamental (FU) and first overtone (FO) pulsators. The 
topology of the instability strip as a function of the chemical composition 
is discussed in \S4. In this section we also address the role of the so-called 
"OR region" to take account of the pulsation properties of mixed-mode pulsators.  
The predicted period-radius (PR) relations and the comparison with similar 
predicted and empirical relations is given in \S5. In \S6 we discuss the 
predicted light curves and their transformation into optical and NIR 
observational planes. The new metallicity dependent PL and PW relations 
are discussed in \S7. The summary of the results of the current investigation 
are outlined in \S8 together with the conclusions and the future developments 
of this project.

\section{Evolutionary and pulsation framework}

\subsection{HB evolutionary models}

The theoretical framework to compute HB evolutionary models has already 
been discussed in detail by~\citet{piet06}. The interested 
reader is referred to the above paper for a thorough discussion concerning 
the input physics adopted for central helium burning evolutionary phases. 
The entire set of HB models are available in the BaSTI 
database\footnote{http://www.oa-teramo.inaf.it/BASTI/}.
The HB evolutionary models were computed, for each assumed chemical 
composition,  using a fixed core mass and envelope chemical profile. 
They were computed evolving a progenitor form the pre-main sequence 
to the tip of the RGB with an age of ~13 Gyr. The RGB progenitor has 
typically a mass of the order of to 0.8 $M_\odot$ in the very 
metal--poor regime increasing up to $\sim$ 1.0 $M_\odot$ 
in the more metal-rich regime.  
The mass distribution of HB models ranges from the mass of the 
progenitors (coolest HB models) down to a total mass of the order 
of 0.5 $M_{\odot}$ (hottest HB models). The trend in He core and envelope mass 
as a function of metal and helium abundances will be discussed in a 
forthcoming paper (Castellani et al. 2015, in preparation).      

The evolutionary phases off the Zero-Age-Horizontal Branch (ZAHB) 
have been extended either to the onset of thermal pulses for more 
massive models or until the luminosity of the model, along the 
white dwarf cooling sequence becomes, for less massive structures, 
fainter than log (L/L$_\odot$)$\sim$-2.5.
The adopted $\alpha$--enhanced chemical mixture is given in Table~1 
of~\citet{piet06}. The $\alpha$-elements were enhanced 
with respect to the~\citet{gn93} solar metal distribution 
by variable factors. We mainly followed elemental abundances for 
field old, low-mass stars by~\citet{ryan91}. The overall 
enhancement---[$\alpha$/Fe]--- is equal to 0.4.

To constrain the metallicity dependence of RRL pulsation properties 
we adopted seven different chemical compositions, namely 
Z= 0.0001, 0.0003, 0.0006, 0.001, 0.004, 0.008 and 0.0198. We also 
assumed, according to recent CMB experiments~\citep{ade14}, 
a primordial He-abundance of 0.245~\citep[][]{cass03}, 
together with a helium--to--metals enrichment ratio of 
$\Delta$Y/$\Delta$Z=1.4. The adopted $\Delta$Y/$\Delta$Z value  
allows us to match the calibrated initial He abundance of the 
Sun at solar metal abundance~\citep{ssb10}.

\begin{figure}
\includegraphics[scale=0.9]{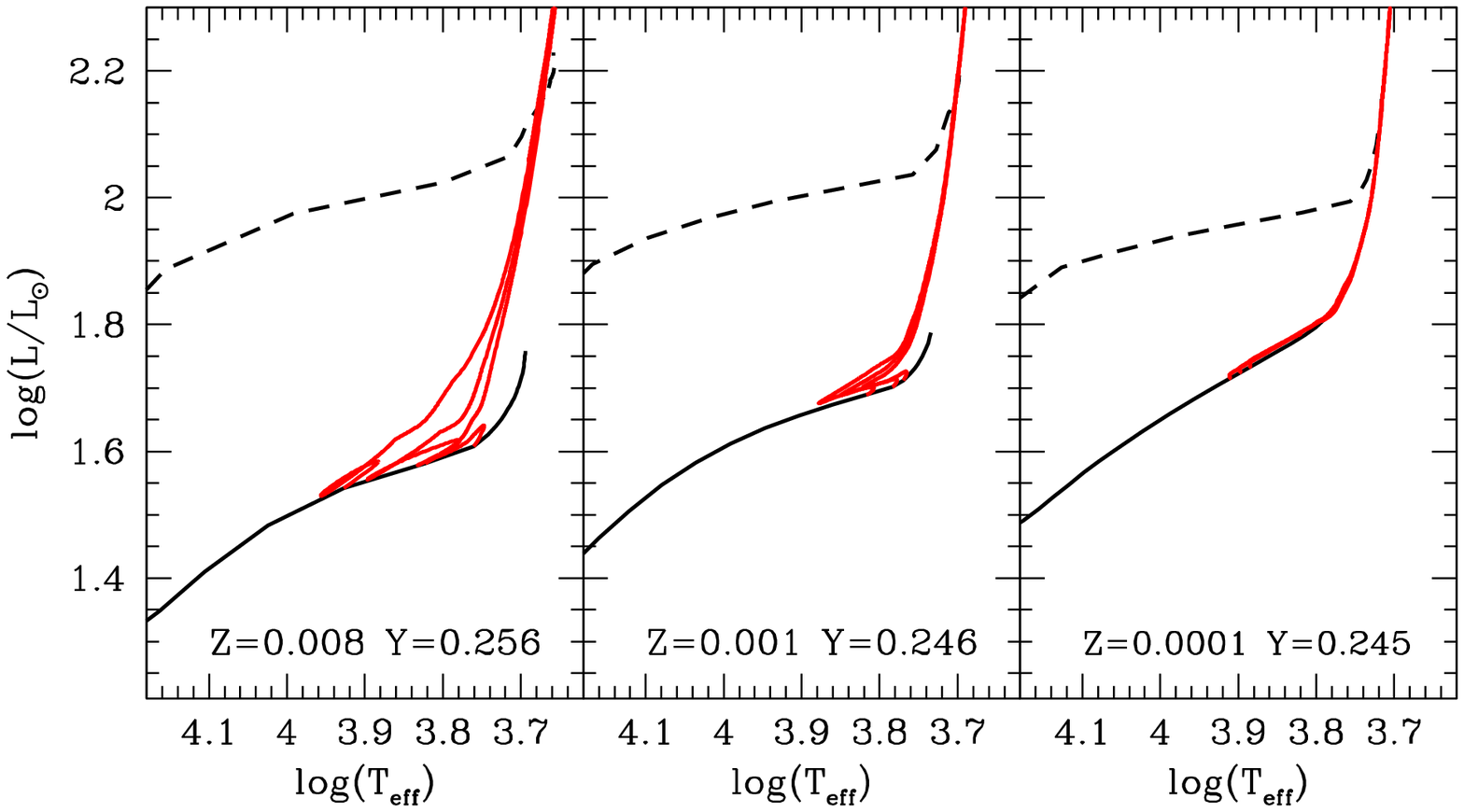}
\caption{The Hertzsprung--Russell diagram for three sets of HB evolutionary 
models. From left to right black solid and dashed lines show the location of 
the ZAHB and of the central helium exhaustion, respectively. Red lines display 
HB evolutionary models populating the RRL instability strip. The individual 
stellar masses are 0.75, 0.76, 0.77 $M/M_\odot$ for the most metal-poor 
(Z=0.0001, right panel), 0.65, 0.66, 0.67 $M/M_\odot$ for the metal--intermediate 
(Z=0.001, middle panel) and 0.56, 0.57, 0.58 $M/M_\odot$ for the more metal--rich 
(Z=0.008, left panel) chemical composition.}\label{fig_hr}
\end{figure}

Figure~\ref{fig_hr} shows the Hertzsprung--Russell diagram for three sets 
of HB models.  From left to right the black solid line shows the 
location of the ZAHB, while the dashed black line the central helium 
exhaustion. The red solid lines display three evolutionary models of 
HB structures populating the RRL instability strip. They range from 
0.75, 0.76, 0.77 $M_\odot$ for the most metal-poor chemical 
composition (right panel), to 0.65, 0.66, 0.67 $M_\odot$ for Z=0.001
(middle panel) and to 
0.56, 0.57, 0.58 for Z=0.008 (left panel).

The above evolutionary prescriptions show three features relevant 
for RRL properties. 

{\em a)}-- An increase in the metal content from Z=0.0001 to Z=0.008    
causes on average, in the middle of the instability strip 
($\log T_{eff}$=3.83--3.85), a decrease of $\sim$0.2 dex in the mean 
luminosity level $\log (L/L_\odot) \sim$1.8 and 1.6), but also a 
decrease in stellar mass (from 0.76 to 0.57 $M/M_\odot$). 
This well known evolutionary evidence~\citep{ccp} and the 
coefficients of both stellar luminosity and stellar mass in the pulsation 
relations (see \S~3 and equations 1) explain the observed decrease in 
period when moving from metal–poor to metal–rich stellar structures 
\citep{fiorentino14}.

{\em b)}-- An increase in the metal content from Z=0.0001 to Z=0.008    
causes a steady increase in the luminosity width of the HB. The 
increase is almost a factor of two in the visual band ($\sim$0.5 
vs $\sim$1 mag). This evidence was originally suggested by~\citet{sandage93} 
and later supported by theoretical models~\citep{bms99}. 
The consequence of this evolutionary feature 
is a steady increase in the intrinsic spread in luminosity during central 
helium burning phases when moving from metal--poor to the metal--rich 
structures.   

{\em c)}-- An increase in the metal content from Z=0.0001 to Z=0.008
causes a steady increase in the width in temperature of the hook 
performed by evolutionary models in the early off--ZAHB evolution. 
The extent in temperature is mainly driven by the efficiency of 
the H-shell burning~\citep{bono97a,cassisi98}. The above 
evolutionary feature affects the lifetime that the different stellar 
structures spend inside the instability strip, and in turn, it opens the 
path to the so-called hysteresis mechanism~\citep{vb73,bccm95a,b95b,bccm97c}. 
To constrain on a more quantitative basis the above effect, we estimated
the typical central He burning time of a metal--poor ($Z=0.0001$, 0.76$M/M_\odot$),
a metal-intermediate ($Z=0.001$, 0.66$M/M_\odot$)  and a more metal-rich
($Z=0.008$, 0.57$M/M_\odot$) stellar structure located in the instability 
strip (see column 4 in Table~\ref{table_time}).  We found that the He-burning lifetime
steadily increases as a function of the metal content from $\sim$67
Myr, to $\sim$77 Myr and to $\sim$86 Myr. This means an increase on average of the order of 25\% when
moving from metal-poor to metal--rich RRLs.
To constrain the difference in
in RRL production rate when moving from a metal--poor to a metal--rich stellar
populations the He-burning lifetimes need to be normalised to a solid evolutionary
clock. We estimate the central H burning phase as the difference between
the main sequence turn-off (MSTO) and a point along the main sequence that is
0.25 dex fainter. To estimate the H-burning lifetimes we selected the stellar
mass at the MSTO of a 13 Gyr cluster isochrone for the three quoted chemical
compositions. The stellar masses and the H-burning lifetimes are also listed
in Table~\ref{table_time}. The ratio between He and H evolutionary lifetimes 
ranges from 0.032 for the metal--poor, to 0.023 for the metal--intermediate to 
0.014 for the metal-rich structures. This means that on average the number of 
RRL per MS star in a metal-poor stellar population is at least a factor of two 
larger than in a metal-rich stellar population. This evidence is further 
suggesting that the number of RRLs and their distribution across the instability 
strip is affected by different parameters: the topology of the instability 
strip, the excursion in temperature of HB evolutionary models and the 
evolutionary lifetimes~\citep{bono97a,m11,fiorentino14}.

 \subsection{Stellar pulsation models}

The pulsation properties of RRLs have been extensively investigated by several 
authors, since the pioneering papers by~\citet{Cox63,Castor71,vanalb71} based 
on linear non adiabatic models.
The advent of nonlinear hydrocodes provided the opportunity to investigate 
the limiting cycle behaviour of radial variables~\citep{chri67,cox74}. 
However, a detailed comparison between theory 
and observations was only possible with hydrocodes taking account of the 
coupling between convection motions and radial displacements. This approach 
provides solid predictions not only on the red edges of the instability 
strip but also on pulsation amplitudes and modal stability~\citep{s82,bs94,feu99,bcm00}.  

Recent developments in the field of nonlinear modelling of RRLs 
were provided by~\citet{szabo04} and by~\citet{smo13}. 
They also take account of evolutionary and pulsational 
properties of RRLs using an amplitude equation formalism.

The key advantage of our approach is that we simultaneously solve 
the hydrodynamical conservations equations together with a 
nonlocal (mixing-length like), time-dependent treatment of 
convection transport~\citep{s82,bs94,bcm00,marconi09}.
We performed extensive and detailed computations of RRL nonlinear 
convective models covering a broad range in stellar masses, 
luminosity levels and chemical composition~\citep[][]{bccm97c,b98,m03,m11}. 
To constrain the impact that the adopted treatment of the convective 
transport has on pulsation observables we have also computed different 
sets of models changing the efficiency of convection~\citep{dmc04,marconi09}.

The RRL pulsation models presented in this paper were constructed using 
the hydrodynamical code developed by~\citet{s82} and updated 
by~\citet[][see also Smolec \& Moskalik 2010 for a similar approach]{bs94,b98,bms99}. 
The physical and numerical assumptions adopted to compute 
these models are the same discussed in~\citet{b98,bms99,m03,m11}.
In particular, we adopted the OPAL radiative opacities released 
by~\citet[][http://www-phys.llnl.gov/Research/OPAL/opal.html]{ir96}
and the molecular opacities by~\citet{af94}.

\begin{figure}
\includegraphics[scale=0.9]{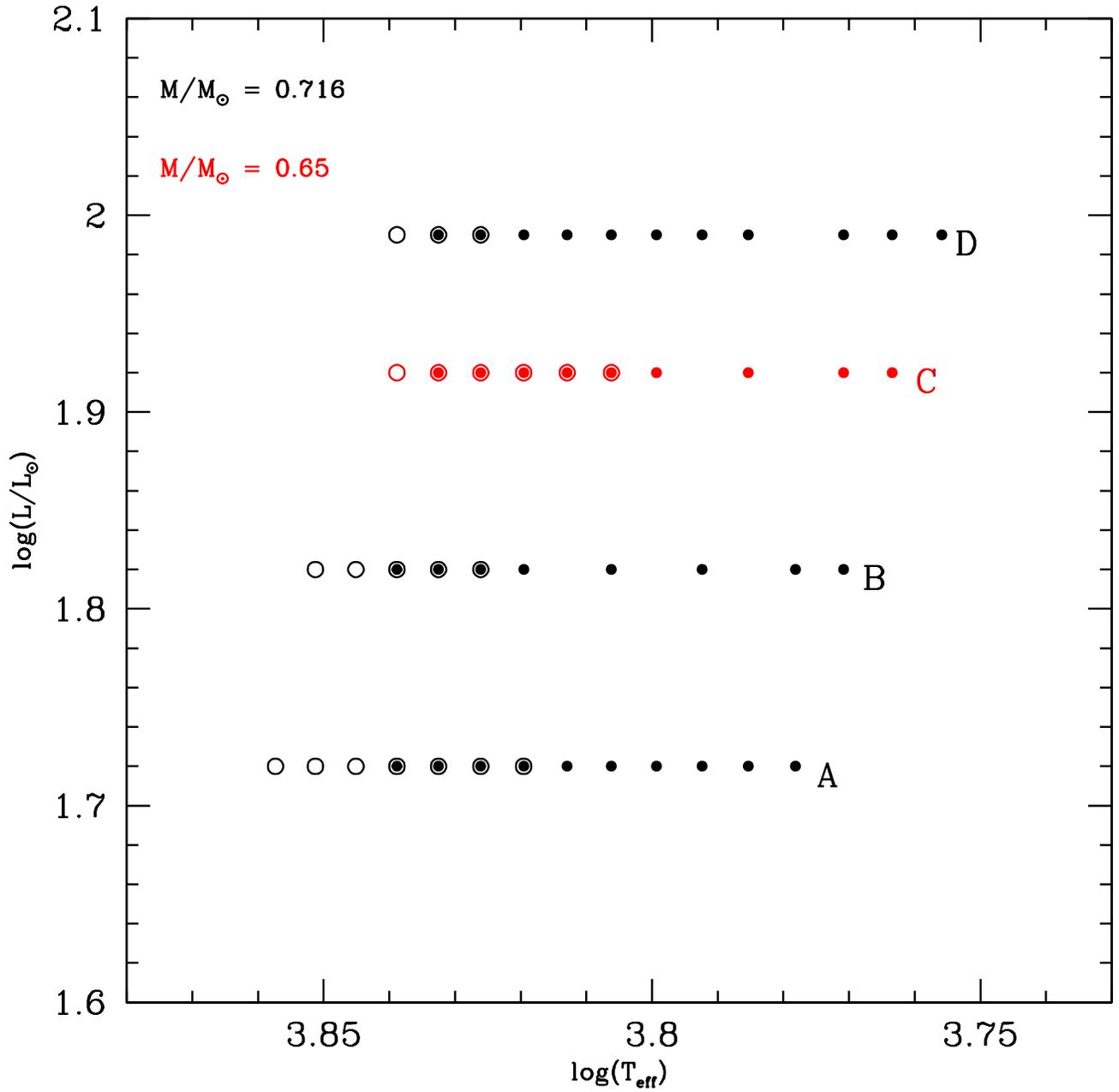} 
\caption{Location in the Hertzsprung--Russell diagram for a set of 
RRL models at fixed chemical composition (Z=0.0003, Y=0.245). 
The FU models are marked with filled circles, while the FOs with 
open circles. The black symbols mark pulsation models computed 
assuming the same stellar mass (0.716 M$_{\odot}$) and three different 
luminosity levels:  the Zero Age Horizontal Branch (ZAHB, sequence A), 
a luminosity level 0.1 dex brighter than the ZAHB (sequence B) and 
the luminosity level of central He exhaustion (sequence D). The red 
symbols display RRL models computed assuming a stellar mass 
$\sim$10\% smaller (0.65 M$_{\odot}$) than the ZAHB mass value 
and 0.2 dex brighter than the ZAHB luminosity (sequence C). This 
sequence of pulsation models was computed to account for evolved 
RRLs. Similar sets of RRL models were computed for the other adopted 
chemical compositions (see Table~\ref{PARAMETRI}).}\label{fig_cmd}
\end{figure}

To compute the new models we adopted seven different metallicities
ranging from the very metal-poor regime ($Z=0.0001$) to the canonical solar
abundance ($Z=0.02$). The helium abundance, for each chemical composition, 
was fixed according to the helium-to-metals enrichment ratio adopted 
by~\citet{piet06}, namely $\Delta{Y}/\Delta{Z} \sim 1.4$ for a primordial 
helium content of  $\sim 0.245$.  
The new RRL models when compared with similar models computed by our group 
present several differences.  
a) The stellar mass, at fixed chemical composition, is the mass of the 
ZAHB ($M_{ZAHB}$) predicted by HB evolutionary models at the center of 
the instability strip ($\log{T_e}\sim3.85$). 
The current predictions indicate that the center of the instability strip is 
located at $\log{T_e}\sim3.82$. We adopted the former value, since this is 
the canonical value and makes more solid the comparison with predictions 
available in the literature. Note that the difference on $M_{ZAHB}$ is minimal 
when moving from $\log{T_e}\sim3.85$ to $\log{T_e}\sim3.82$.        
b) The faintest luminosity level, at fixed chemical composition, is 
the predicted ZAHB luminosity level. 
c) To take account of the intrinsic width in magnitude~\citep{sand90} 
of the RRL region we also adopt, at fixed chemical composition and 
ZAHB mass value, a luminosity level that is 0.1 dex brighter than the 
ZAHB luminosity level.  
d) Theory and observations indicate that stellar structures located close 
to the blue edge of the instability strip cross the instability strip 
during their off ZAHB evolution. These stellar structures typically evolve 
from hotter to cooler effective temperatures and cross the instability 
strip at luminosity levels higher than typical RRL~\citep[$\omega$ Cen,][]{m11}.   
To take account for these evolved RRLs we also adopted, at fixed chemical 
composition, a second value of the stellar mass---$M_{evo}$---that is 
10$\%$ smaller than $M_{ZAHB}$. The decrease in the stellar mass was estimated 
as a rough mean decrease in stellar mass, over the entire metallicity range, 
between the ZAHB structures located close to the blue edge and at the 
center of the instability strip. The luminosity level of $M_{evo}$ was 
assumed equal to 0.2 dex higher than the ZAHB luminosity level. Once again 
the assumed luminosity level is a rough estimate of the increase in luminosity 
typical of evolved RRLs over the entire metallicity range. 
To provide the physical structure and the linear eigenfunctions adopted 
by the hydrodynamical models, we computed at fixed chemical composition, 
stellar mass and luminosity level a sequence of radiative hydrostatic 
envelope models. The physical assumptions adopted to construct the linear 
models are summarized in Appendix 1, together with the linear blue
edges based on linear hydrostatic models.
The mean absolute bolometric magnitude and effective temperature of the 
pulsation models approaching a stable nonlinear limit cycle is evaluated as 
an average in time over the entire pulsation cycle. This means that they are 
slightly different when compared with the static initial values, i.e. with the 
values the objects would have in case they were not variables. The difference 
is correlated with the luminosity amplitude and the shape of the light curve 
of the variables. The reader interested in a more detailed discussion is 
referred to \citet{bono95d}. The difference in 
luminosity for the models located outside the instability strip is negligle. 
The difference in effective temperature is negligle for models hotter than 
the blue edge and marginal for those cooler than the red edge.
 In Figure~\ref{fig_cmd} we show the location in the Hertzsprung--Russell diagram 
of a set of RRL models at fixed chemical composition (Z=0.0003, Y=0.245). 
The FU models are marked with filled circles, while the FOs with 
open circles. The black symbols mark pulsation models computed 
assuming the same stellar mass (0.716 M$_{\odot}$) and three different 
luminosity levels: the ZAHB (sequence A), a luminosity level 0.1 dex 
brighter than the ZAHB (sequence B) and the luminosity level of central 
He exhaustion (sequence D). The red symbols display RRL models computed 
assuming a stellar mass $\sim$10\% smaller (0.65 M$_{\odot}$) than 
the $M_{ZAHB}$ mass value  and 0.2 dex brighter than the 
ZAHB luminosity level (sequence C). This sequence of pulsation models 
was computed to take account of evolved RRLs.

The adopted stellar parameters of the computed pulsation models
are listed in Table~\ref{PARAMETRI}. In the first three columns are 
given the metallicity, the helium content and the stellar mass on the 
ZAHB at the center of instability strip (M$_{ZAHB}$). The fourth column 
lists the three selected luminosity levels. For each adopted chemical 
composition, stellar mass, luminosity level and effective temperature, 
we investigated the limit cycle stability of RRL models both in the 
FU and FO mode.
The nonlinear pulsation equations were integrated in time till the
limit cycle stability of radial motions approached their asymptotic
behavior, thus providing robust constraints not only on the boundaries of the 
instability strip (IS), but also on the pulsation amplitudes.  

\section{New metal-dependent pulsation relations}

The correlation between pulsation and evolutionary observables is rooted 
in several analytical relations predicting either the absolute magnitude 
(PLC relations) or the pulsation period (pulsation relation) as a function 
of stellar intrinsic parameters (Bono et al. 2015, in preparation). 
The pulsation relation and its dependence on stellar mass, stellar luminosity 
and effective temperature was cast in the form currently adopted more than 
40 years ago by~\citep{vanalb71}. To derive the so-called van Albada \& Baker
(vAB) relation they adopted linear, nonadiabatic, convective models.  
The radial pulsation models are envelope models, i.e. they neglect the 
innermost and hottest regions of stellar structures. This means that they 
neglect nuclear reactions taking place in the center of the stars and assume 
constant luminosity at the base of the envelope. The base of the envelope 
is typically fixed in regions deep enough to include a significant fraction 
of their envelope mass, but shallow enough to avoid temperatures hotter 
than $\sim 10^7$ K. This means that the computation of pulsation model
does require the knowledge of the mass-luminosity relation of the stellar 
structures we are dealing with. van Albada \& Baker  in their seminal 
investigations adopted evolutionary prescriptions for HB stars by~\citet{ir70}. 
       
Modern versions of the vAB relation have been derived using updated 
evolutionary models and/or nonlinear pulsation models~\citep{cox74}. 
More recently, they have also been derived using nonlinear convective 
models and including a metallicity term~\citep[see e.g.][]{bccm97c,m03,dmc04}
to take account of the metallicity dependence of the ZAHB luminosity level. 
In view of the current comprehensive approach in constructing RRL pulsation 
models, we computed two new vAB relations for FU and FO pulsators. We found  

\begin{eqnarray}
 \log{P_F}&=&(11.347\pm0.006)+(0.860\pm0.003)\log{L/L_{\odot}}+\,
 \nonumber \\
&-&(0.58\pm0.02)\log{M/M_{\odot}}-(3.43\pm0.01)\log{T_e}+ \, \nonumber \\
&+&(0.024\pm0.002)\log{Z} \,
\end{eqnarray}

\begin{eqnarray}
 \log{P_{FO}}&=&(11.167\pm0.002)+(0.822\pm0.004)\log{L/L_{\odot}}+\,
 \nonumber \\
&-&(0.56\pm0.02)\log{M/M_{\odot}}-(3.40\pm0.03)\log{T_e}+ \, \nonumber
\\
&+&(0.013\pm0.002)\log{Z} \,
\end{eqnarray}

where the symbols have their usual meaning. The standard deviation 
for FU pulsators is 0.06 dex, while for FO pulsators is 0.002 dex.
The decrease in the intrinsic spread for FO pulsators was expected, 
since the width in temperature of the region in which FOs attain a stable nonlinear limit cycle is  on average  $60\%$ of the one  for FU pulsators. 

\begin{figure}
\includegraphics[scale=0.9]{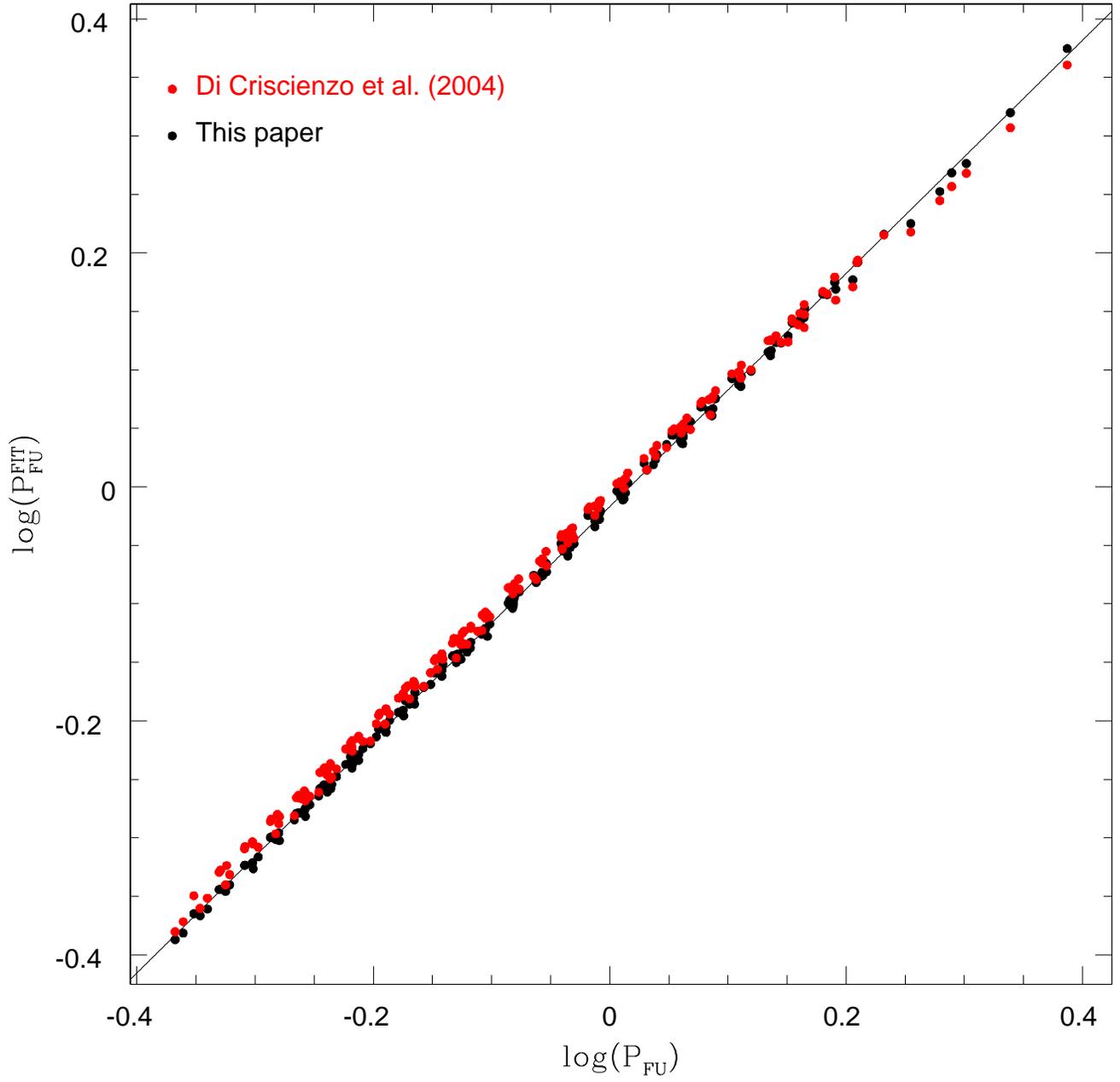}
\caption{Comparison between the FU periods ($\log(P_{FU})$) predicted by 
individual models and the FU periods ($\log(P^{FIT}_{FU})$) given by the 
pulsation relation provided in this paper (black symbols). The red symbols 
display the same comparison, but using the pulsation relation provided 
by~\citet{dmc04}.}\label{compper}
\end{figure}

Figure~\ref{compper} shows the comparison between the computed FU periods 
and the ones predicted by the above pulsation relation 
(black symbols) and the same comparison but adopting the relation 
by~\citet{dmc04} (red symbols).
We notice that the differences between the quoted pulsation relations 
are within the intrinsic scatter of the linear regressions. 

When dealing with cluster variables, one of the typical approaches to improve the sample size is to fundamentalize the periods of FO 
pulsators. The classical relation adopted for the fundamentalization 
is $\log{P_{FU}}=\log{P_{FO}}+ 0.127$ 
~\citep[][and Coppola et al. 2015, in preparation]{dmc04,braga14}. 
However, this relation is based on a very limited sample of observed 
double-mode pulsators~\citep[][]{p91} and its applicability to RRL 
has never been verified. 

\begin{figure}
\includegraphics[scale=0.9]{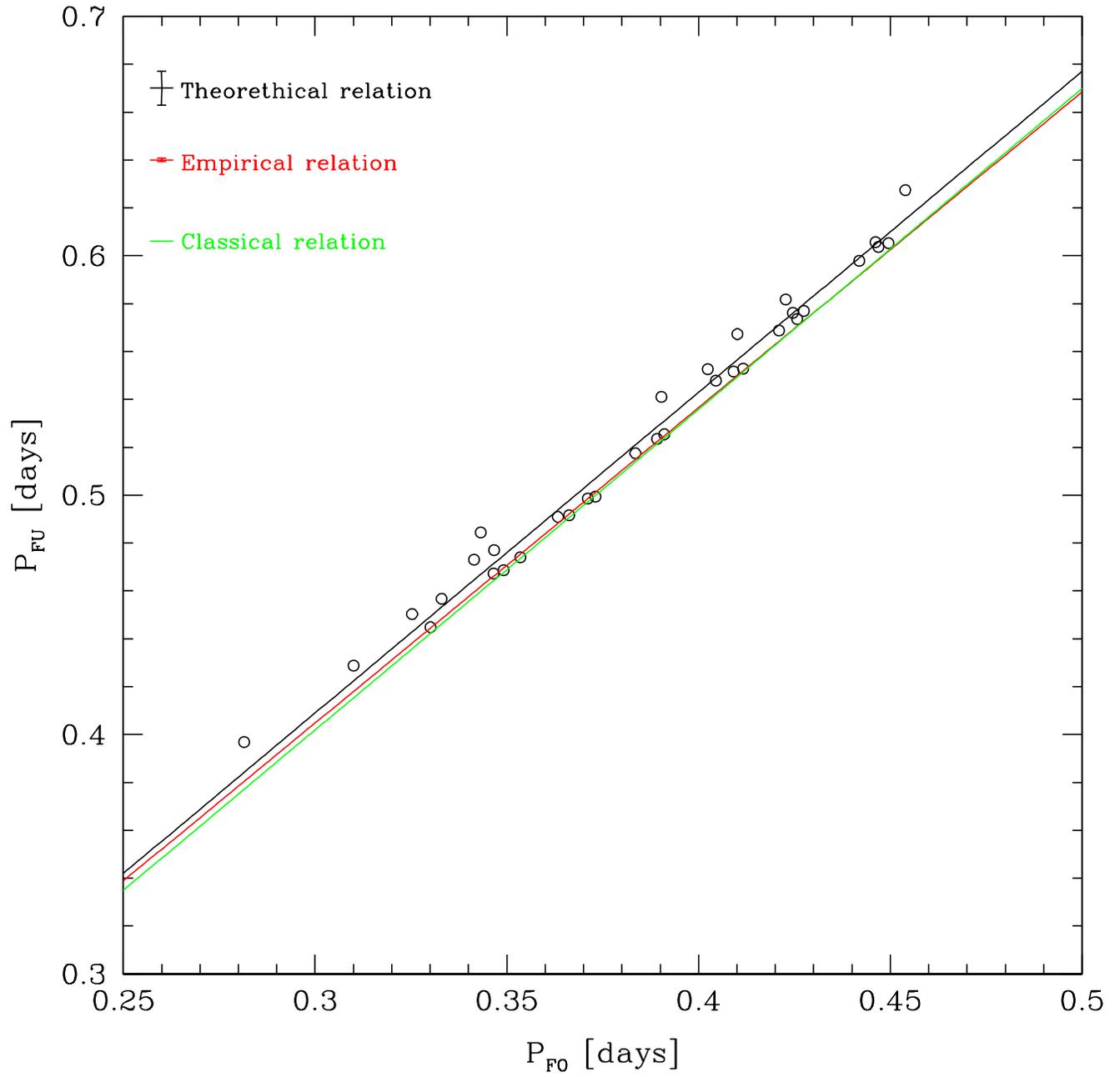}
\caption{Relation between FU and FO periods for pulsation models located inside 
the so-called "OR region". The black solid line shows the linear regression 
based on theoretical models. The red and the green solid lines display the 
empirical and the classical relation between FU and FO periods (see text 
for more details).}\label{fundamentalized}
\end{figure}

Figure~\ref{fundamentalized} shows the relation between 
FO and FU period for models located inside the so called ”OR region” of the instability strip: the region located 
between the FU blue edge and the FO red edge (see below), where the two modes 
approach pulsationally stable nonlinear limit cycles. The ensuing 
linear relation (black solid line) is compared with the empirical 
relation (red solid line) obtained by using a sample of $\sim$ 80
known double-mode variables identified in different stellar 
systems (Galactic globulars, dwarf spheroidals) and available 
in the literature (Coppola et al. 2015, in preparation). 
The classical relation is also shown for comparison 
(green solid line). Data plotted in this figure indicate that the 
new theoretical relation (black line) attains FU periods that are, 
at fixed FO period, slightly longer than the empirical ones 
(red line). On the other hand, the classical fix (green line) 
attains FU periods that are, at fixed FO period, slightly shorter 
than the empirical ones. We conclude that the comparison shown in 
Figure~\ref{fundamentalized} indicates that the classical relation 
is a very plausible fix over a broad range of metal abundances.

\section{Topology of the instability strip}

The current theoretical framework allows us to predict the approach to 
nonlinear limit cycle stability of the different pulsation modes. This 
implies the opportunity to constrain the topology of the instability 
strip, i.e. the regions of the Hertzsprung-Russell diagram in which 
the radial modes approach a pulsationally stable nonlinear limit cycle. 
The anonymous referee noted that for the FU/FO models the approach to limit 
cycle stability for a nonlinear system would imply strictly periodic
oscillations. However, in the current theoretical framework the approach to a 
nonlinear limit cycle stability also means small changes in the mean magnitude 
and effective temperature (see \S 2.2).
An original approach to compute exact periodic solutions of the nonlinear 
radiative pulsation equations was presented by \citet{Ste74,Ste83}. 
However, we still lack a similar relaxation scheme--Flouquet analysis--for 
radial oscillations taking account of a time-dependent convective transport 
equation.
A similar, but independent approach was also developed by \citet{bg84} using amplitude equation formalism, i.e. the temporal evolution of 
modal amplitudes are described by a set of ordinary differential equations. 
Canonical amplitude equations, including cubic terms, were derived to 
investigate radiative models, but nonlinear convective models required 
the inclusion of quintic terms \citep{bu99}. However, their  
calculation using static models is not trivial, therefore,
\citet{Kol98}, \citet{szabo04} and  \citet{SmoMo08a} decided to 
couple the solution of nonlinear conservation equations with amplitude 
equation formalism. The nonlinear limit cycle stability based on this 
approach is very promising, but it is very time consuming, since 
radial motions have to be analysed over many pulsation cycles.

However, in the current context, we define that a radial mode approaches 
a pulsationally stable nonlinear limit cycle when period and amplitudes over 
consecutive cycles attain their asymptotic behavior. {\em Stricto sensu} they 
are not exactly periodic, but {\em lato sensu} they approach a periodic 
behavior. Therefore, following ~\citet{bono93} and~\citet{bcm00}, the limit cycle in a nonlinear time-dependent convective 
regime was considered pulsationally stable when the differences in the pulsation 
properties over consecutive cycles become smaller than one part per thousands. 
This means that we are integrating the entire set of equations for a number 
of cycles ranging from a few hundreds to several hundreds.     

\begin{figure}
\includegraphics[scale=0.9]{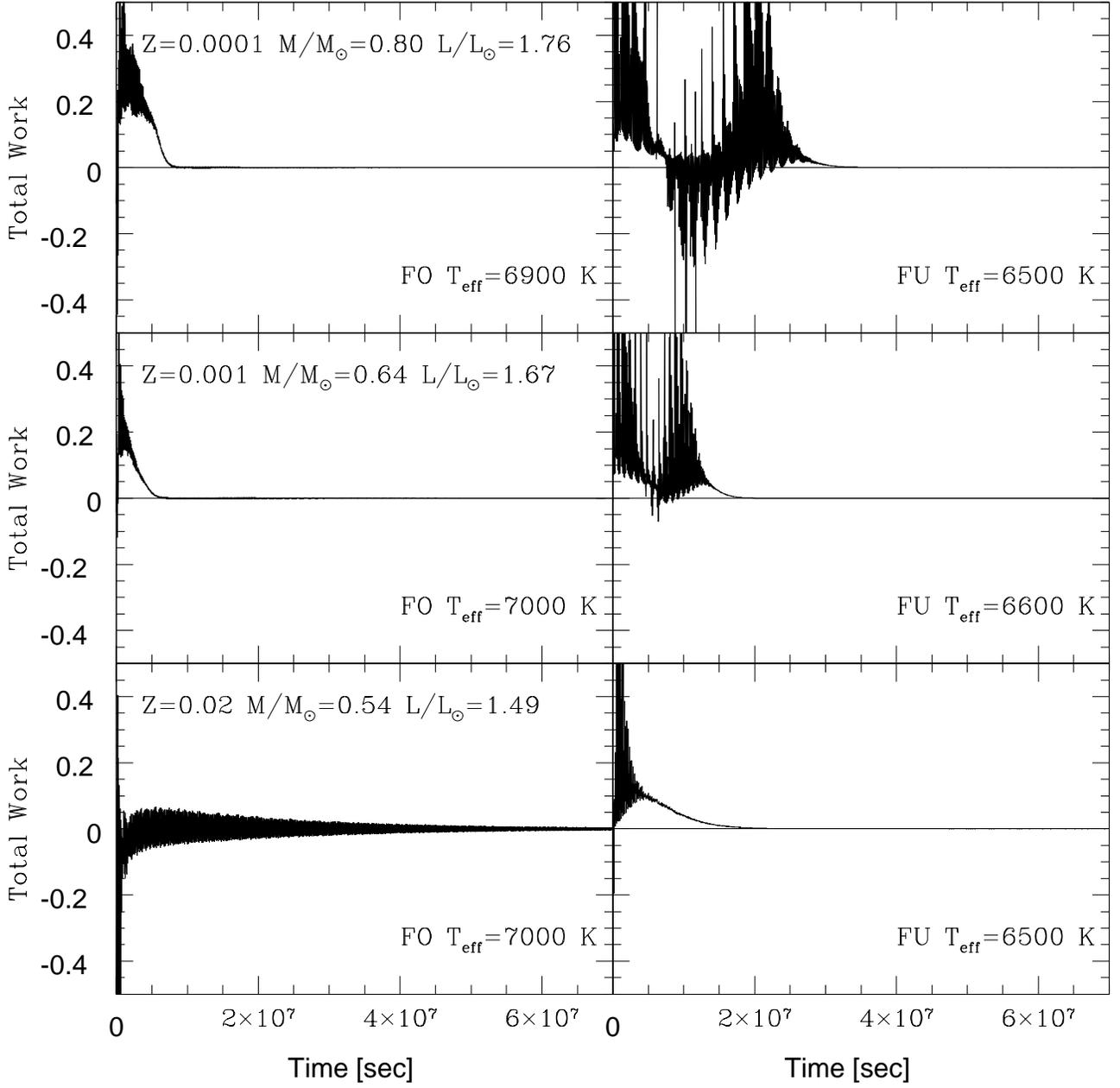}
\caption{Nonlinear total work integral as a function of integration time for 
three different FO (left panels) and three FU models (right panels). They
are centrally located in the middle of the instability strip, and 
constructed assuming three different metal abundances, stellar masses 
and luminosity levels (see labeled values).} \label{work}
\end{figure}

\begin{figure}
\includegraphics[scale=0.9]{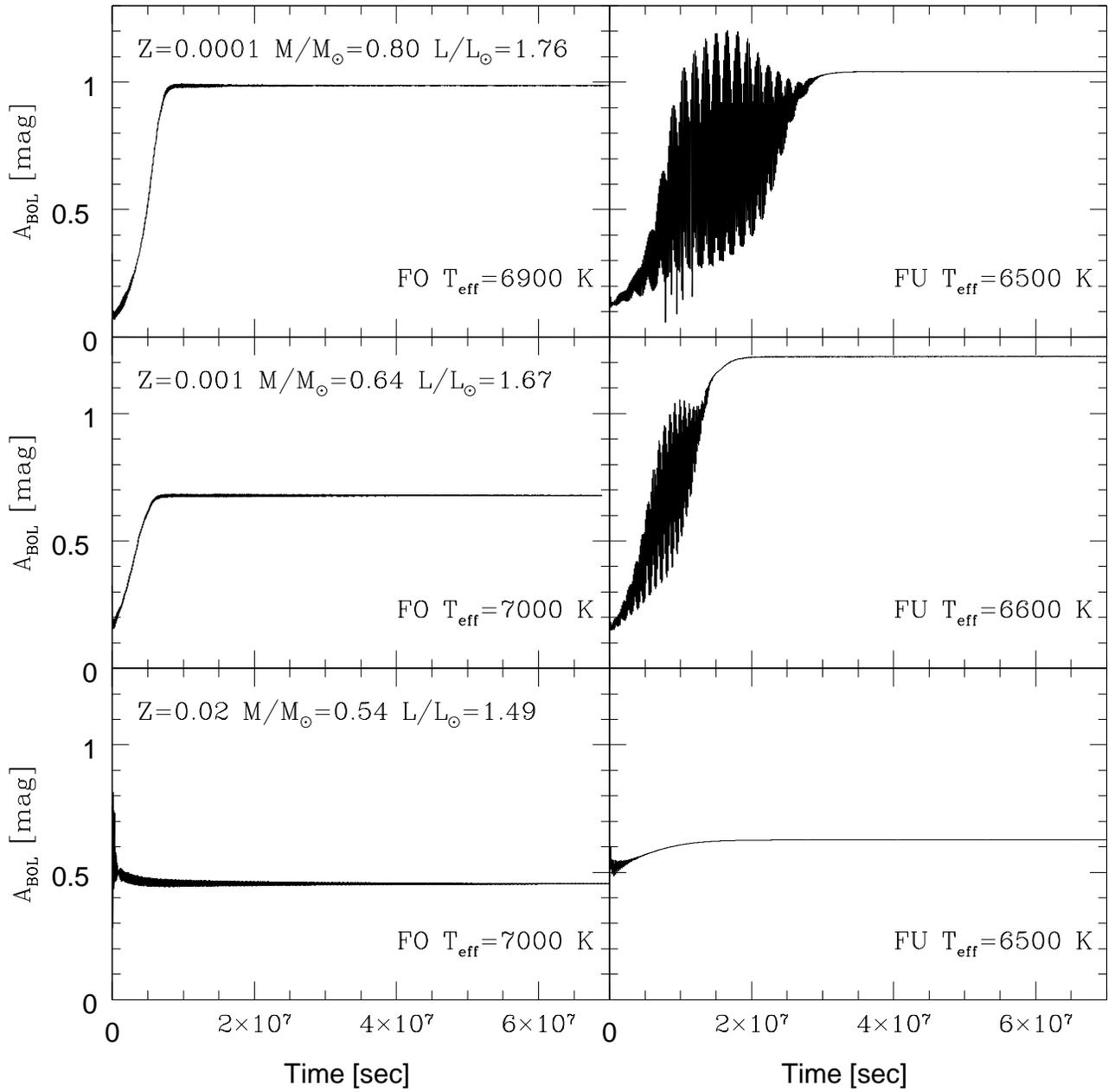}
\caption{bolometric amplitude as a function of the integration 
time for the same models of Fig. ~\ref{work}. }\label{amp}
\end{figure}

To support on a more quantitative basis the above definition, Fig.~5 shows 
the nonlinear total work integral as a function of integration time for 
three different FO (left panels) and FU models (right panels). They are 
centrally located in the middle of the instability strip, and constructed 
assuming three different metal abundances, stellar masses and luminosity 
levels (see labeled values). 
Theoretical predictions plotted in this figure display that the dynamical 
behavior, after the initial perturbation\footnote{The nonlinear analysis 
was performed by imposing a constant velocity amplitude of 10 km/s both to 
FU and FO linear radial eigenfunctions (see Appendix).}, approaches the 
pulsationally stable nonlinear limit cycle. The transient phase is at most 
of the order of $\sim$ 100 cycles (top left model), and indeed the relative 
changes in total work after this phase are smaller than $\pm$0.0001. 
To further constrain the approach to a pulsationally stable nonlinear limit 
cycle, Fig.~\ref{amp} shows the bolometric amplitude as a function of the integration 
time for the same models of Fig. ~\ref{work}. Once again the luminosity amplitudes 
approach their asymptotic behavior  after the transition phase. After this 
phase the relative changes in bolometric amplitudes over consecutive cycles 
are smaller than $\pm$0.001 mag.

 To constrain the location of the boundaries of the instability strip 
we adopted, for each fixed chemical composition, a step in effective temperature 
of 100 K. The sampling in temperature of the pulsation models becomes slightly 
coarser across the instability strip. The effective temperatures of the 
edges of the instability strip for 
FU and FO pulsators and for the adopted chemical compositions are listed 
in Table~\ref{PARAMETRI}. The blue (red) edges are defined, for each 
sequence of models,  50 K hotter (cooler) than the first (last) pulsating model 
in the specific pulsation mode.

Figure~\ref{strip} shows the boundaries for FU (solid lines) and FO (dashed lines) 
for three selected chemical compositions, namely $Z=0.0001$ (top panel), $Z=0.001$ 
(middle panel) and $Z=0.02$ (bottom panel). The boundaries of the instability 
strip, when moving from the hot to the cool region of the HR diagram, are the 
FO blue edge (FOBE), the FU blue edge (FBE), the FO red edge (FORE) and the 
FU red edge (FRE).  The region of the instability strip located between the 
FBE and the FORE is the so called "OR region". In this region the RRLs 
could pulsate simultaneously in the FU and in the FO mode.
Note that we still lack a detailed knowledge of the physical mechanisms that 
drive the occurrence of mixed-mode pulsators~\citep{b96}. This means 
that we still lack {\em ab initio} hydrodynamical calculations approaching 
a pulsationally stable nonlinear double-mode limit cycle. The occurrence of 
a narrow region of the instability strip in which double-mode pulsators attain 
a stable nonlinear limit cycle was suggested several years ago by \citet{szabo04}. However, the approach adopted by these authors to deal 
with the turbulent source function in convectively stable regions was criticized 
by \citet{SmoMo08a,SmoMo08b}. The occurrence of double-mode pulsators has  
been investigated among classical Cepheids \citep{SmoMo08b} and 
$\delta$ Scuti/SX Phoenicis stars \citep{bono02a}. However, the modelling 
of these interesting objects is still an open problem that needs to be addressed       
on a more quantitative basis.

It is empirically well known that mixed-mode RRLs are 
located in a defined region of the instability strip, between the long period tail 
of FO pulsators and the short period tail of FU pulsators (Coppola et al. 2015, 
in preparation). The above observational scenario is further supporting 
theoretical predictions~\citep[][]{b97b} suggesting the occurrence of mixed-mode 
pulsators in a narrow range in effective temperatures of the instability strip. 

The instability strips plotted in Figure~\ref{strip} display 
several interesting features worth being discussed in more detail. 

{\em i)}-- The increase in the metal content causes a shift of the instability 
strip towards cooler (redder) effective temperatures. This evidence was 
originally suggested by~\citet[][]{b97b} and supported by empirical
evidence~\citep{stetson14,fiorentino14}.
Note that the current prediction 
for the FOREs in the metal-rich regime, need to be cautiously treated. There 
is evidence that current prediction are slightly redder than suggested by 
empirical evidence~\citep{bccm97c}.   

{\em ii)}-- The instability strip becomes, when moving from metal--poor 
to metal--rich pulsators, systematically fainter. Moreover, the range 
in luminosities covered by the three selected luminosity levels increases 
when moving from metal-poor to metal--rich stellar structures. The above 
evidence is a direct consequence of HB evolutionary
properties~\citep{piet04,piet06}, originally brought forward on an 
empirical basis by~\citep{sand90}.    

{\em iii)}--  The region of the instability strip in which the FOs approach a pulsationally 
stable nonlinear limit cycle vanishes at higher luminosity levels. The 
FOBE and the FORE tend to approach the same effective temperature.
This point of the instability strip was originally called 
"intersection point"~\citep{s75} and suggests 
the lack of long period FO pulsators~\citep{bs94,bccm97c}. This evidence 
is soundly supported by observations~\citep{k13}. In passing, we note 
that the current predictions and similar calculations available in the 
literature~\citep{bs94,bccm97c} do suggest the possible presence of 
stability isles, i.e. regions located at luminosities higher than the 
intersection point in which the FOs attain a pulsationally stable nonlinear limit cycle. More 
detailed theoretical and empirical investigations are required to 
constrain the plausibility of the above predictions.    

{\em iv)}-- The width in temperature of the instability strip in which 
 FO pulsators attain a pulsationally stable nonlinear limit cycle becomes systematically narrower 
when moving from metal-poor to metal-rich pulsators. The predicted trend 
appears quite clear, but we still lack firm empirical constraints. Note 
that the metal-rich regime is not covered by cluster variables, since 
the most metal-rich globulars hosting RRLs are NGC~6388 and NGC~6441 
and both of them are more metal--poor than Z$\sim$0.006. The empirical 
scenario is still hampered by the lack of wide spectroscopic surveys  
of Halo and Bulge RRLs. 
 
{\em v)}-- We note that the extreme edges of the strip, namely the FOBE 
and the FRE follow a linear behaviour and this occurrence is true for 
all the assumed metal contents. On the basis of this evidence, we derived 
the following analytical relations for the FOBE (rms = 0.003) and the 
FRE (rms= 0.006) as a function of the assumed metallicity.

\begin{eqnarray}
 \log{{T_e}^{FOBE}}&=&(-0.080\pm0.008)\log{L/L_{\odot}}+ \, \nonumber \\
&-&(0.012\pm0.002)\log{Z}+3.957\pm0.003 \,
\end{eqnarray}

\begin{eqnarray}
 \log{{T_e}^{FRE}}&=&(-0.084\pm0.009)\log{L/L_{\odot}}+\, \nonumber \\
&-&(0.012\pm0.002)\log{Z}+3.879\pm0.006\,
\end{eqnarray}

Note that the above relations suggest that the width in effective 
temperature of the entire instability strip, among the different 
chemical compositions, is constant and of the order of 1300 K.

Figure~\ref{strip_conf} shows the comparison between these
relations and the edges (FOBE, FRE) previously determined 
by~\citet{dmc04} for the three labelled metal contents.
The agreement is good and the differences are within $\pm 50 K$ in
effective temperature, that is half of the step in temperature adopted 
in the quoted grids of models.

\subsection{Pulsation properties inside the {\it OR region}}

 We already mentioned in Section 4, that the modelling of radial pulsators that 
oscillate simultaneously in the FU and in the FO (double-mode) is still an open 
problem. In this context we define "OR region" the region of the instability 
strip in which pulsation models, after the initial perturbation, attain a 
pulsationally stable nonlinear limit cycle either in the FU or in the FO. 
This means that the approach to the nonlinear limit cycle does depend on the 
adopted initial conditions (linear radial eigenfunctions).
From the theoretical point of view the models located in the OR region were 
adopted to mimic the properties of double-mode pulsators. The same objects are 
called, from the observational point of view, RRd-type variables.
The RRd variables play a fundamental role in constraining 
the evolutionary and the pulsation properties of RRLs. Indeed, 
the so-called Petersen diagram~\citep[][]{p91,b96,bra01}, 
i.e. the period ratio between FO and FU periods 
($P_{FO}/P_{FU}$) versus the FU periods ($P_{FU}$), is a good 
diagnostic to constrain the pulsation mass and the intrinsic
luminosity of RRd variables.      
The key advantage in using the Petersen diagram is that the adopted 
observables are independent of uncertainties affecting the distance 
and the reddening correction of individual objects. Moreover and 
even more importantly, theory and observations indicate that both 
the mean magnitudes and colors of the two modes are within the errors 
the same~\citep[][]{sosz14}. The above evidence and 
the pulsation relations discussed in \S3 indicate that the Petersen 
diagram can be soundly adopted to constrain the actual mass of 
RRd variables.   

A detailed theoretical investigation of double-mode RR Lyrae in the Petersen 
diagram was provided by \citet{pdc00}. They 
investigated in detail the dependence of the period ratio on intrinsic 
parameters (stellar mass, luminosity, effective temperature, metal abundance). 
Moreover, they also performed a detailed comparison with RRd in several globular 
clusters and nearby dwarf galaxies. In particular, they found that Large 
Magellanic Cloud double mode variables cover a modest range in metallicity 
(-1.7$\le$[Fe/H]$\le$-1.3). Moreover, the spread in period ratio at fixed iron 
abundance could be explained as a spread in stellar mass. Theoretical \citep{b96,kw99,k00} and empirical 
\citep{beau97,alc99,sosz11,sosz14} evidence 
indicates that more massive RRd variables attain, at fixed FU period, larger 
period ratios.

Moreover, theory and observations suggest that metal-rich 
RRd variables show, when compared with metal-poor objects 
shorter FU periods and smaller period 
§ratios~\citep[][]{bra01,dicrisci11,sosz14}.

To take account of both the dependence on the stellar mass 
and on the metallicity, we computed the period ratios of the 
pulsation models located in the "OR region" for each assumed 
chemical composition. Figure~\ref{fig_rel_mass} shows the 
distribution of the quoted models in a 3D logarithmic plot 
including the period ratio ($P_{FO}/P_{FU}$), the stellar mass 
in solar units and the metal content. Data plotted in this 
figure display a well defined correlation. Therefore, we performed 
a Least Squares linear regression and we found the following 
analytical relation:

\begin{eqnarray}
log M/M_\odot &=& -0.85 (\pm 0.05) -2.8 (\pm 0.3) log (P_{FO}/P_{FU}) +\,
\nonumber \\
&-&0.097 (\pm0.003) log Z  \,
\end{eqnarray}

where the symbols have their usual meaning. The above relation, with a
rms of 0.004,  
provides the unique opportunity to constrain the pulsation mass 
of double-mode pulsators on the basis of their period ratios 
and metal-contents (Coppola et al. 2015, in preparation), observables that are independent of uncertainties affecting 
individual distances and reddenings.

\section{The Period-Radius relation}

Recent improvements in interferometric measurements~\citep{k08,ertel14}
provided the opportunity to measure the diameter of several evolved radial 
variables (classical Cepheids, Kervella et al. 2001; Mira, Milan-Gabet 2005). 
The next generation of optical and NIR 
interferometers~\citep[][]{n00,g14} will allow us to measure the diameter 
of nearby field RRLs. Moreover, recent advancements in the application of 
the Infrared Surface Brightness (IRSB) method to RRLs is providing new and 
accurate measurements of RRL mean radii.  

Therefore, we decided to use the current sequences of nonlinear, convective 
models to constrain the Period-Radius (PR) relation of FU and FO RRLs.  
Figure~\ref{pr_f} shows in a logarithmic plane the mean radius of FU pulsators 
over the entire set of chemical compositions versus the FU period. We performed a 
linear Least Squares regression over the entire set of models and we found 
the following PR relation for FU pulsators (black line):

\begin{equation}
\log R/R_{\odot} = 0.866(\pm 0.003) +0.55(\pm 0.02) \log P
\end{equation}

with a standard deviation of 0.03 dex. The red and the green solid lines display 
the predicted PR relation by~\citet{marconi05}, based on similar RRL models 
and the extrapolation to shorter periods of the predicted PR relation
of BL Herculis variables,  
provided by~\citet{m07} (MDC07), using the same theoretical framework adopted 
in this investigation. The standard deviations of the individual PR relations 
plotted in the top left corner of the same figure indicate a good agreement 
among the different predicted PR relations over the entire period range. 

The difference between the new PR relation and the one derived 
by~\citep{marconi05} is the consequence of the different assumptions 
concerning stellar masses and luminosities adopted for the different  
chemical compositions. In the current approach, we adopted 
evolutionary prescriptions, while in~\citep{marconi05} the grid of 
pulsation models was constructed by adopting, for each chemical 
composition and stellar mass, a fixed step in luminosity level. 
In spite of the different approach adopted in selecting the evolutionary 
parameters, the agreement is quite good. The difference becomes slightly 
larger only in the BL Herculis regime, i.e. for $\log P >0$.  
 
Note that the period range adopted in the above figure 
is larger than the typical range of RRL stars ($\sim 0.2 \le P \le 1.0$ days, 
Marconi et al. 2011). To validate the above theoretical scenario    
Figure~\ref{pr_f} also shows the comparison with the empirical PR relation 
provided by~\citet{bm86} (dashed green line) for both RRL and BL Herculis variables. 
The agreement is once again quite good over the entire period range.

The similarity of both predicted and empirical PR relations for RRL and 
BL Herculis supports earlier suggestions concerning the tight evolutionary 
and pulsation correlation of these two classes of evolved low-mass radial 
variables. This applies not only to the PR relations~\citep{bm86,m07}, 
but also to the PL relation~\citep{caputo04,matsu09,rip15}.   
In this context, the BL Herculis are just the evolved component of RRL stars,
i.e. HB stellar structures that in their off-ZAHB evolution cross the instability 
strip at higher luminosity levels than typical RRLs~\citep{m11}. 
In spite of this indisputable similarity between RRL and BL Herculis a word of caution 
is required. According to the above evolutionary framework the BL Herculis evolve 
from the hot to the cool side of the instability stripe. This means that they 
require a good sample of hot HB stars. However the HB morphology, i.e. the 
distribution of HB stars along the ZAHB, does depend on the
metallicity. The larger 
is the metallicity, the redder becomes the HB morphology~\citep{castellani83,renzini83}. 
This means that the probability to produce BL Herculis becomes vanishing in 
metal-rich stellar systems. On the 
other hand, we have evidence of RRL stars with iron abundances that are either 
solar or even super-solar in the Galactic Bulge~\citep{wt91}. This would imply 
that the above similarity between RRL and BL Herculis variables might 
not be extended over the entire metallicity range.

Predicted mean radii plotted in Figure~\ref{pr_f} display, at fixed period, 
a large intrinsic dispersion. To further constrain this effect and to estimate    
the sensitivity of the PR relation on the metal content, we also computed     
new Period-Radius-Metallicity (PRZ) relations for both FU and FO pulsators. 
We found  

\begin{eqnarray}
\log R/R_{\odot}&=&0.749(\pm 0.006) +0.52(\pm 0.03) \log P +\,
\nonumber \\
 &-&0.039(\pm0.006) \log Z \,
\end{eqnarray}

with a standard deviation of 0.006, for FU models and

\begin{eqnarray}
\log R/R_{\odot} &=& 0.87(\pm 0.03) +0.60(\pm 0.06) \log P +\,
\nonumber \\
&-&0.033(\pm0.005) \log Z\,
\end{eqnarray}

with a standard deviation of 0.006, for FU models and 0.003 for FO models.

The models and the new PR relations are plotted in the top (FU) and in the 
bottom (FO) panel of Figure~\ref{pr_f_fo}. Again, similar relations 
by~\citet{marconi05} are shown for comparison. As expected the PRZ relations 
display more tight correlations than the classical PR relation. Thus suggesting 
a clear dependence on the metal content. The above evidence indicates that the 
PRZ relation is a powerful tool to constrain individual radius estimates for RRL 
of known period and metallicity with a precision of the order of 1\%.

\section{Predicted light curves}

One of the key advantages in using nonlinear, convective hydrodynamical models 
of RRL stars is the possibility to predict the variation of the leading observables 
along the pulsation cycle. Among them the bolometric light curves play a fundamental 
role, since mean magnitudes and colors do depend on their precision. In the current 
investigation, we computed limit cycle stability for both FU and FO RRLs covering a 
broad range in chemical compositions and intrinsic parameters (stellar mass, luminosity 
levels). The different sequences of models do provide an atlas of bolometric light 
curves that can be adopted to constrain not only the mean pulsation properties of RRLs, 
but also the occurrence of secondary features (bumps, dips) along the pulsation cycle.  
The Figures~\ref{fig_cdl1} and~\ref{fig_cdl2} show predicted FU and FO bolometric 
light curves for the sequence of models at $Z=0.0006$, $Y=0.245$, $M=0.67M_{\odot}$, 
$\log{L/L_{\odot}}=1.69$\footnote{The bolometric light curves for the other chemical 
compositions are available upon request to the authors.}.

\subsection{Transformation into the observational plane}

The bolometric light curves discussed above have been transformed into the most 
popular optical (UBVRI) and NIR (JHK) bands using static model 
atmospheres~\citep[][]{bono95d,cast97a,cast97b}. 

Once the bolometric light curves have been transformed into optical 
and NIR bands, the magnitude-averaged and intensity-averaged mean
magnitudes and colors can be derived for the entire set of stable pulsation 
models.  The intensity-averaged magnitudes are the most used in the literature, 
since they overcome different thorny problems connected with the shape 
of the light curves~\citep{bono93}\footnote{The magnitude-averaged 
quantities are available upon request to the authors.}. They are 
presented in the following.

\subsection{Mean magnitudes and colors} 

The Tables~\ref{medieintF} and~\ref{medieintFO} give the intensity-averaged 
mean magnitudes for the entire grid of FU and FO models. For each set of models 
the Tables list the chemical composition (metal and helium content) together 
with the intrinsic parameters (stellar masses and luminosity levels) adopted 
for the different sequences. The first three columns list, for every pulsation 
model, the effective temperature, the logarithmic luminosity level
and the logarithmic period. The subsequent eight columns give the predicted 
Johnson-Kron-Cousins ($UBVRI$) and NIR ($JHK$, 2MASS photometric system) 
intensity-averaged mean magnitudes.

When moving the instability strip from the HR diagram into the observational planes, 
the predicted FOBE and FRE colors can be correlated with the absolute magnitude 
and the metal content according to the relations given in
Table~\ref{bordicol}. We note that in the case of $K-(J-K)$ the
predicted boundaries do not depend on metallicity.

\section{Metal-dependent Period-Luminosity and Period-Wesenheit relations}

The current grid of nonlinear, convective RRL models provide the unique 
opportunity to build a new and independent set of metal-dependent diagnostics 
to determine RRL distances. We focussed our attention on metal--dependent 
PL (PLZ) relations and on the metal--dependent PW relations (PWZ).  
This approach was adopted for the entire set of optical, optical-NIR and NIR 
intensity-averaged mean magnitudes.

Note that to derive the PLZ and the PWZ relations for FU pulsators 
we included, for each assumed chemical composition, sequences A,B 
and C of models (see Table 2 and Figure 2). We neglected the sequence 
D models, since these luminosity levels are too bright for the bulk 
of RRLs. They are more typical of radial variables at the transition 
between BL Herculis and W Virginis variables.    

To derive the PLZ and the PWZ relations for FO pulsators we included 
the two faintest luminosity levels (sequences A and B). We neglected the 
brighter sequences C and D, since they are typically brighter than the 
"Intersection Point" (see \S4).

 \subsection{The PLZ relations}

The predicted PLZ relations were derived for FU and FO pulsators covering 
the entire set of chemical compositions (Z=0.0001--0.02). We performed 
several linear Least Squares relations of the form:   

\begin{equation}
M_X = a\;+\;b \log P\;+\;c [Fe/H]
\end{equation}

Together with the independent PLZ relations for FU and FO pulsators we also estimated 
an independent set of "global" PLZ relations using simultaneously FU and FO pulsators. 
The periods of the latter group were fundamentalized using the classical relation 
(see also \S3).  This is the typical approach adopted to improve the sample size, 
and in turn, the precision of cluster/galaxy distances based on
RRLs~\citep[][Coppola et al. 2015, in preparation]{braga14}.    
The classical relation is $\log P_{FU}=\log P_{FO}+0.127$ and the symbols 
have their usual meaning. 

Theoretical~\citep{bono01,catelan04} and empirical~\citep{benko11,braga14} 
evidence indicates that RRLs do not 
obey to a well defined PL relation in the blue ($UBV$) bands. The 
slope of the PL relation becomes more and more positive when moving 
from the $R$ to the NIR bands. The slope is the consequence of a 
significant variation in the bolometric corrections when moving 
from the blue (short periods) to the red (long periods) edge of the 
instability strip. This change is minimal in the blue bands and 
becomes larger than one magnitude in the NIR bands~\citep{b03}.   
These are the reasons why we derived new PLZ relations, only for the 
$RIJHK$ bands. The coefficients of the $RIJHK$ PLZ relations are listed 
in Table~\ref{tab_plz1} together with their uncertainties and standard 
deviations. 

The left panels of Figure~\ref{pl1} show the PLZ relations for FU and 
FO pulsators, while the right ones the global PLZ relations. Lines of 
different color display relations with metal contents ranging from 
$Z=0.0001$ (black) to $Z=0.02$ (purple). The current models soundly 
support the empirical evidence~\citep[][]{benko11,coppola11} 
that the slope of the PL relations becomes steeper when moving from the optical to the 
NIR bands. Indeed, the slope increases from 1.39 in the $R$-band  to 2.27 
in the $K$-band. Moreover, the standard deviation of the PLZ relations decreases by a 
factor of three when moving from the optical to the NIR bands. This means that the 
precision of individual distances, at fixed photometric error,  increases in the 
latter regime. The above trends are mainly caused by the temperature dependence 
of the bolometric correction as a function of the wavelength. 

Moreover, optical and NIR absolute magnitudes of metal-poor RRLs 
are, at fixed period, systematically brighter than metal-rich ones. The difference
is mainly due to evolutionary effects. The ZAHB luminosity at the effective 
temperature typical of RRLs ($\log T_e$=3.85) decreases for an increase in 
metal content~\citep{pietrinferni2013}. Note that the metallicity dependence 
is, within the errors, similar when moving from the optical to the NIR 
bands. Indeed the coefficient of the metallicity term attains values of 
the order of 0.14--0.18 dex.  

Figure~\ref{pl_conf} shows the comparison between predicted and empirical 
$K$-band PL relations available in the literature. The top panel shows 
the comparison for the most metal--poor chemical composition ($Z=0.0001$), 
while the bottom panel for a metal-intermediate chemical composition 
($Z=0.001$). Note that in the comparison we only included PL relations 
taking account either of a metallicity term~\citep{bono01,catelan04} or 
of the HB morphology~\citep{cassisi04} or based on an empirical 
approach~\citep{b03,s06}. 

It is worth mentioning that the PLZ relation 
provided by~\citet{b03} is based on field and cluster RRLs for which the 
individual distances were estimated by using the NIR surface-brightness 
method~\citep[][]{storm1994}. On the other hand, the PLZ relation provided 
by~\citet{s06} was estimated using RRLs from 16 calibrating GCs to fix the 
slope of the relation and RR Lyr itself to fix the zero-point of the relation. 
The above comparison indicates that the current $K$-band PL relations agree 
quite well with similar predicted~\citep{catelan04} (blue lines) and 
empirical~\citep{s06} (cyan lines) PL relations. The difference is on average 
smaller than 1$\sigma$ both for metal-poor and metal-intermediate chemical 
compositions.  

The current predictions attain intermediate slopes when compared with similar 
predictions by~\citet{bono01} (red lines) and by~\citet{b03} (green lines).   
The differences are caused either by different assumptions concerning the 
ML relations adopted to construct the grid of pulsation models~\citep{bono01}  
or by different assumptions concerning the absolute calibrators and/or the 
metallicity scale~\citep{b03}. 

The above discussion is mainly focussed on the comparison between predicted 
and empirical slopes of the $K$-band PL relations. A glance at the PL relations 
plotted in top panel of Figure~\ref{pl_conf} indicates that the zero-point 
of the current metal--poor 
prediction is $\sim$0.1 mag brighter than the empirical PL relation by~\citet{s06}.   
However, the difference becomes marginal in the metal-intermediate regime (bottom panel). 
The difference might be the consequence of the adopted zero-point. The empirical 
$K$-band PL relation is rooted on RR Lyr itself, which is a 
metal-intermediate ([Fe/H]=-1.50$\pm$0.13 dex; Braga et al. 2014) field RRL.

\subsection{Dual band PWZ relations}
The PLZ relations have several advantages in estimating individual distances. 
However, they are prone to uncertainties affecting reddening corrections. 
This problem is strongly mitigated when moving into the NIR and MIR
regime~\citep{madore12}, but still present. The problem becomes even more severe
in regions affected by differential reddening. To overcome this problem it 
was suggested $\sim$40 years ago~\citep[][]{vb75,madore82} to use reddening 
free pseudo-magnitudes called "Wesenheit magnitudes". They are defined as:

\begin{equation}
W(X,Y) = M_X + \xi (M_X-M_Y)
\end{equation}

where the coefficient of the color term -- $\xi$ -- is the ratio between the selective 
absorption in the X band and the color excess in the adopted color. It is 
fixed according to an assumed reddening law that in our case is 
the~\citet{cardelli} law. The coefficients of the color terms are
listed in the third column of Table~\ref{tab_pwz}.  

Pros and cons of both optical and NIR PW relations have been widely discussed 
in the literature~\citep{inno13}. However, they have been typically 
focussed on classical 
Cepheids~\citep[][]{c00b,f02,f07,marconi05,nk05,b10,storm11,ripepi12}. 
We briefly summarise the key 
issues connected with optical and NIR PW relations. The reader is referred 
to~\citet{braga14} and to~\citet{inno13} for a more detailed 
discussion. The pros are the following:  
a) Individual distances are independent of reddening uncertainties. 
b) The PW relation mimic a PLC relation, since they 
include a color term. This means that individual distances are more 
precise when compared with distances based on PL relations. 
c) The metallicity dependence for classical Cepheids appears to be 
vanishing for optical-NIR and NIR PW relations~\citep{c00b,marconi05,inno13}.     
The main cons are the following: 
a) The PW relation assume that the reddening law is universal. However, 
the use of bands with a large difference in central wavelength mitigates 
the problem~\citep{inno13}.  
b) Two accurate mean magnitudes are required to provide the distance. 
   
To investigate the properties of optical and NIR PW in the RRL regime 
we computed new FU, FO and global PW relations for the different combinations 
of the above five bands. They were derived using the entire metallicity 
range (Z=0.0001--0.02). The coefficients, their errors and standard 
deviations are listed in Table~\ref{tab_pwz}.
The panels of Figures~\ref{pw} shows from top to bottom the PW relations 
based on optical bands. The color coding of the different lines is the same 
as in Figure~\ref{pl1}. 
A glance at the predictions plotted in this figure discloses an 
interesting result. The PW(V,B-V) relations plotted in the top panels 
show a minimal, if any, dependence on the metal content. The coefficients 
of the metallicity term for FU, FO and global PWZ relations are, within 
the current uncertainties, similar and vanishing (Table~\ref{tab_pwz}).  
This finding is at odds with similar PWZ relations for classical 
Cepheids. The metallicity dependence for these objects is larger 
in the optical regime and becomes smaller either in the NIR or in the 
optical-NIR regime. The difference is mainly due to the higher range 
in surface gravities and effective temperatures covered by RRLs when 
compared with classical Cepheids.  It is worth mentioning that the 
above finding is even more appealing if we account for the fact 
that RRLs cover a range in metallicity that is at least one dex 
wider than the range covered by classical Cepheids 
([Fe/H]$\simeq -2.5$ -- $+0.30$  vs [Fe/H]$\simeq -1.5$ -- $+0.30$;
~Pedicelli et al. 2009; Romaniello et al. 2008; Dambis et al. 2014).

The above findings bring forward two important consequences: 
a) The PW(V,B-V) relation are robust distance indicators for RRL 
sharing simultaneously the advantages to be reddening free and 
almost independent of the metal content.  
b) The use of the  PW(V,B-V) relation together with independent 
optical and NIR magnitudes can also provide tight constraints on 
individual RRL reddenings. The main cons is, once again, the 
dependence on the assumed reddening law. 

To constrain this effect we adopted two independent 
reddening laws, namely the ones by~\citet{mccall04} and~\citet{fitz09}.
We found that the color coefficient of the PW(V,B-V) relation 
change from -3.07~\citep{mccall04} to -3.36~\citep{fitz09}. This 
difference causes a difference in distance of the order of 0.2\%. 

The PW relations plotted in Figures~\ref{pw} and listed in Table~\ref{tab_pwz} 
indicate that the coefficient of the metallicity term becomes of the order of 
0.1 in the other optical bands, while it increases to 0.15--0.20 in the 
optical-NIR and in the NIR bands. Thus suggesting a similar dependence of the 
optical-NIR and NIR regimes on the metal content. 

However, the standard deviation has a different behavior. It is on average a 
factor of two smaller in the NIR-bands that in the optical ones. 
The smallest values are attained by optical-NIR PW relations 
in which they are of the order of a few hundredths of magnitude. 
The above evidence indicate that ultimate precision of individual 
distances is a balance between photometric and spectroscopic 
uncertainties. Thus suggesting that the error budget is the main 
criterium in the selection of the appropriate PWZ relation to estimate 
individual RRL distances.

To validate the above PWZ relations we compared the current predictions 
with similar mass-dependent PW relations available in the literature. 
In particular, we focussed our attention on the BV and the VI filter 
combinations provided by~\citet{dmc04}. Note that the latter predictions 
rely on models covering a narrower metallicity range ($Z \le 0.006$).
Figure~\ref{plw_conf} shows the comparison for the BV filters in two 
different metal-abundances. For each selected metal abundance we 
adopted the stellar mass predicted by HB evolutionary models for 
a ZAHB structure located at $\log{T_e}=3.85$ (see Table~\ref{medieintF}). 
These mass values and the the mass-dependent relations 
provided by~\citet{dmc04} were adopted for the two labelled metallicities.
The agreement is excellent for $Z=0.001$ (right panel) and within 
1$\sigma$  for the more metal-poor, $Z=0.0001$ (left panel), 
chemical composition. The difference is mainly caused by a difference 
in the zero-point, and in turn, in the adopted ML relation. The slopes 
attain both in the metal-poor and in the metal-rich regime similar values. 

The comparison of the $VI$ filters is of little use, since the coefficient of 
the color term adopted by~\citet{dmc04} is slightly different than the current 
one (1.433 vs 1.54)\footnote{The difference is caused by a different assumption 
concerning the central wavelength of the Johnson-Kron-Cousins $I$-band adopted to 
estimate the selective absorption from the reddening law~\citep{cardelli}. In particular,~\citet{dmc04} adopted 
$\lambda_{I}=7940$ \AA, while we adopted  $\lambda_{I}=8060$ \AA~\citep{bessel90}.}.  
To overcome the problem we adopted their coefficient and we found that both 
the zero--points and slopes agree quite well in the metal-poor and in the 
metal--rich regime.

\subsection{Triple band PWZ relations}

The dual band PW relations have been widely used in the recent literature in 
dealing with RRL and Cepheid individual distances. During the last few years 
it has been suggested by~\citet{riess11}, on an empirical basis, to use triple 
bands PW relations. These new diagnostics can be defined as  

\begin{equation}
W(X,Y,W) = M_X + \xi (M_Y-M_W)
\end{equation}

where the coefficient of the color term --$\xi$-- is the ratio between the 
selective absorption in the X band and the color excess in the adopted 
$(Y-W)$ color.
In the quoted paper the authors adopted the $F555W$ and the $F814W$ bands 
plus the NIR $F160W$ band in the WFC3 HST photometric system and they applied 
the new PWZ relation applied to Classical Cepheids in external galaxies. 
The same approach, but based on nonlinear, convective Cepheid models was 
also adopted by~\citet{fiorentino13}.   
More recently, triple bands PW relations have been derived by~\citet{braga14} 
for optical, optical-NIR and NIR PW relations of RRLs in M4. 

According to the quoted authors, the key advantages in using triple bands 
PW relations to estimate individual distances are the following: 
{\em i)} they have smaller intrinsic dispersions when compared with dual 
band PW relations; 
{\em ii)} they are less prone to possible systematics in using the same 
magnitude in the color term.  
The main drawback is the need of accurate mean magnitudes in three 
different bands.

Here we derive for the first time triple band PWZ relations using 
RRL pulsation models. They are listed in Table~\ref{tab_pwz_3bands} 
and shown in Figure~\ref{plw_3bands} for different combinations.
Note that we focussed our attention on triple PWZ relations including 
a NIR magnitude and an optical color. We found that the coefficients 
of the color term are typically smaller than 0.7 and even smaller than 
0.2 for colors including bands with large differences in central 
wavelengths ($B-R$, $B-I$). The coefficients of the metallicity term 
attain on average values similar to the dual bands PWZ relations.
The standard deviations of the quoted triple bands PWZ relations are, 
once again, similar to the standard deviations of dual bands PWZ relations.

\subsection{Metal-independent PWZ relations}

We found that a few dual and triple band PWZ relations listed in
  Tables~\ref{tab_pwz} and \ref{tab_pwz_3bands}, have coefficients of the metallicity term that are smaller than 0.05 mag/dex. 
This implies a vanishing dependence on the metal content, since a difference 
of 1 dex in metal content would imply a difference in distance modulus at most 
of the order of 0.05~mag. Therefore, we decided to compute for the same 
PWZ relations and new set of PW relations that neglect the metallicity 
dependence. The zero-points and the slopes of the metal-independent relations 
are listed in Table~\ref{PWZindep}. The new PW relations show standard 
deviations that are only slightly larger than the metal-dependent PWZ relation. 
Thus further supporting the marginal role played by the metal content over the 
entire period range. This theoretical evidence, once validated on an empirical 
basis, opens the path to a new diagnostic to estimate individual distances of 
RR Lyrae for which the metal content is not available.

\subsection{Uncertainties affecting the coefficients of PLZ and PWZ relations}

The new theoretical framework for RR Lyrae stars we are developing does depend 
on the physical assumptions adopted in the treatment of turbulent convection. 
We use a nonlinear, nonlocal, time-dependent approach to deal with convective 
transport. However this treatment, and similar approaches available in the 
literature \citep{SmoMo08a,smolec10}, do rely on a free parameter, 
the so-called mixing length parameter. In the current approach, we adopted 
a mixing length parameter equal to 1.5. Plausible changes of this parameter 
mainly affect the pulsation amplitudes and to a minor extent the boundaries 
of the instability strip. Detailed calculations concerning the dependence of 
nonlinear observables on the efficiency of the convective transport have 
already been discussed by \citet{dmc04}. The treatment of the 
turbulent convection also relies on the use of a few other free parameters. 
They have been fixed according to the value of the mixing length parameter 
following the prescriptions provided by\citet{bs94}. The adopted values have been 
validated fitting the light curves of field and cluster RR Lyrae \citep{bcm00,mc05,md07}, Bump Cepheids \citep{bono2002c}, the prototype $\delta$ Cephei \citep{nmb08} 
and classical Cepheids in eclipsing binaries \citep{m13}. Moreover, 
we do allow the convective flux to attain negative values at the boundaries 
of convective stable regions \citep{bcm00}. A similar 
approach was also adopted by \citep{SmoMo08a}.  

Bolometric light curves are transformed into the observational plane using 
bolometric corrections and color-temperature (CT) relations predicted by 
static stellar atmosphere models. This means that we are assuming a quasi-static 
approximation in transforming  predicted observables 
(static vs effective surface gravity). A proper treatment does require 
the detailed solution of the radiative transfer equation \citep{df99}. In spite of the 
limitations of the current approach, we found that transformations based 
on  an independent set of stellar atmosphere models
\citep[PHOENIX][]{ku06} do provide very similar results. In this context, it is worth mentioning that a detailed comparisons between 
predicted and observed mean effective temperatures of RR Lyrae stars indicate 
a difference of the order of 150 K \citep{cacciari00}. Moreover, we still lack 
a detailed comparison between color-temperature relations based on static and 
hydrodynamical atmosphere models.

In the current investigation we adopted, following \citet{piet06}, 
a $\alpha$-enhanced chemical mixture. Moreover, we are adopting, following \citet{cass03}, a primordial helium abundance of $Y_P$=0.245, 
consistent with measurements of the  cosmic microwave background \citep{pry02,planck14,hi13} and a helium-to-metal enrichment ratio of 1.4. 
Larger helium abundances, at fixed metal content, affect the pulsation 
properties of RR Lyrae \citep{bono95c,m11}. A systematic 
investigation of the impact that helium abundance has on pulsation observables 
will be investigated in a forthcoming paper.

Finally, the evolutionary predictions adopted in this investigation are also 
affected by uncertainties in the adopted input physics \citep{cassisi98,cassisi07} 
and on the treatment of mass loss, atomic diffusion, extra-mixing and neutrino 
losses. The reader interested in a more detailed analysis of the uncertainties 
affecting the predicted ZAHB luminosity is referred to
\citet{cassisi98}, \citet{salaris13}, \citet{valle13}, \citet{van13}.

\section{Summary and future remarks}

We present a comprehensive theoretical investigation of the pulsation 
properties of RRL stars. To provide a homogeneous and detailed 
theoretical framework to be compared with the huge photometric 
and spectroscopic data sets that are becoming available in the 
literature we compute a large grid of nonlinear, convective hydrodynamical 
models of RRL stars. The RRL models were constructed assuming a broad range 
of metal abundances ($Z=0.0001$--$0.02$) and at fixed helium-to-metals 
enrichment ratio ($\Delta{Y}/\Delta{Z}$=1.4). 
 As a whole we computed $\approx$420 nonlinear hydrodynamical models. Among them 
$\approx$300 display a pulsationally stable nonlinear limit cycle, while 
$\approx$60 experience a mode switching. The latter group approaches, after a 
transient phase, a nonlinear limit cycle that is different from the initial 
perturbed linear radial eigenfunction. Moreover, $\approx$60 models quench 
radial oscillations because they are located outside the instability strip. 
They are either hotter than the FOBE or cooler than the FRE.
The main difference of the current grid when compared with similar calculations 
available in the literature is that for each fixed chemical composition, 
the stellar mass and the luminosity levels adopted to construct pulsation 
models were fixed according to detailed central He burning HB evolutionary 
models. In particular, for each fixed chemical composition we adopted two 
different stellar masses to take account of RRL located either in the 
proximity of the ZAHB (sequence A in Figure 2) or crossing the instability 
strip, from hot to cool effective temperatures, at higher luminosity levels 
(sequence C in Figure 2). Indeed the former models 
were constructed by assuming three different luminosity levels (A, B, D) to 
take account of the off-ZAHB evolution till central He exhaustion.     

To provide a thorough analysis of the topology of the RRL instability strip, 
as a function of the metal content we computed the pulsation stability for 
both FU and FO pulsators. The calculations were extended in time till the 
individual models approached limit cycle stability and we could constrain 
their modal stability. The main results of the above theoretical framework 
are the following.  
  
$\bullet$ {\em Pulsation properties.} The current theoretical framework allowed 
us to provide detailed predictions of a broad range of observables (amplitudes: 
luminosity, velocity radius, effective temperature; mean magnitudes, velocities, 
radii). In particular, we investigated their dependence on the metal content.    

$\bullet$ {\em Modal stability.} We provided a detailed mapping of the RRL instability 
strip as a function of the metal content. We found that an increase in metal content 
causes a systematic shift of the instability strip towards redder (cooler) colors. 
This confirms previous findings by our group.   

$\bullet$ {\em Pulsation relations.} We provided accurate pulsation relations 
(van Albada \& Baker relations) for both FU and FO pulsators. The key advantage 
of the current relations is that they rely on a homogenous evolutionary and pulsation
framework.    

$\bullet$ {\em Pulsation masses of double mode pulsators.} The topology of the 
instability strip allowed us to constrain the so-called "OR" region, i.e. 
the region in which both FU and FO pulsators attain a pulsationally stable nonlinear limit cycle.
Models located in this region were adopted to mimic the properties of 
double mode pulsators. We derived a new analytical relations to constrain 
the mass of double mode pulsators using period ratios and metal content.  

$\bullet$ {\em Period radius relations.} We derived new Period-Radius-Metallicity 
(PRZ) relations for FU and FO pulsators. They agree quite with previous PR and 
PRZ relations.  

$\bullet$ {\em Transformations into the observational plane.} Bolometric magnitudes 
and effective temperatures were transformed into the observational plane using 
bolometric corrections and color-temperature relations provided
by~\citet{cast97a,cast97b}. 
This means that we provide intensity-averaged mean magnitudes 
and colors together with luminosity amplitudes in the most popular optical 
and NIR bands ($UBVRIJHK$). 

$\bullet$ {\em RR Lyrae as distance indicators.} Homogeneous predictions concerning 
optical/NIR mean magnitudes allowed us to compute new diagnostics to estimate 
individual distances of Galactic and Local Group RRL stars.  

{\tt Period-Luminosity-Metallicity relations --} We derived new accurate 
PLZ relations for RRL stars in optical/NIR bands (RIJHK). We confirm that RRL stars 
do not obey to PLZ relations in the blue regime. The lack of a well defined slope 
is the minimal dependence of the bolometric correction, in these bands, on the 
effective temperature. 

{\tt Period-Wesenheit-Metallicity relations --} We derived new accurate 
PWZ relations for RRL stars in optical, optical-NIR and NIR bands. The key advantages 
of the PWZ relations is that they are independent of reddening uncertainties, moreover, 
they also show smaller intrinsic dispersion when compared with similar PLZ relations. 
The latter feature is the consequence of the inclusion of a color term mimicking 
a PLC relation. The main drawback is that they rely on the 
assumption that the reddening law is universal. However, theoretical and empirical 
evidence indicate that optical/NIR PWZ relations are less prone to secondary features 
of the reddening law. To fully exploit the use of the PWZ relations as distance 
indicators we computed both dual and triple band PWZ relations. The latter appear 
very promising, but they do require three independent mean magnitudes. 
 Finally, we found that the predicted PW(V,B-V) relations are almost
 independent of the metal content.

The current investigation is the first step of a large project aimed at constraining 
the pulsation properties of RRL as a function of chemical composition and ages of 
the progenitors. We plan to investigate the dependence on the helium content in a 
forthcoming investigation. Moreover, we also plan to transform the current 
predictions from the UV to the MIR using homogeneous sets of stellar atmosphere 
models. This new theoretical scenario shall pave the road for a massive use of RRL stars 
as distance indicators. The above plan appears even more compelling in waiting
for the first Gaia data release together with the advent of new space (JWST) and
ground--based observing facilities. In this context the Extremely Large Telescopes
(European-ELT [E-ELT]\footnote{http://www.eso.org/public/teles-instr/e-elt.html},
the Thirty Meter Telescope [TMT]\footnote{http://www.tmt.org/} and the Giant
Magellan Telescope [GMT]\footnote{http://www.gmto.org/})
will play a crucial role, since they are going to resolve individual HB stars
in Local Group and in Local Volume galaxies~\citep{bsg13}.

\acknowledgments
It is a pleasure to acknowledge the anonymous referee for his/her positive words concerning our investigations and for her/his pertinent comments and suggestions that improved the content and the cut of the current manuscript. This work was partially supported by PRIN-INAF 2011 "Tracing the formation 
and evolution of the Galactic halo with VST" (P.I.: M. Marconi), by PRIN-INAF 
2012 "The M4 Core Project with Hubble Space Telescope" (P.I.: L. Bedin) and by 
PRIN-MIUR (2010LY5N2T) "Chemical and dynamical evolution of the Milky
Way and Local Group galaxies" (P.I.: F. Matteucci). One of us (G.B.) thanks 
The Carnegie Observatories visitor programme for support as science visitor.

 {}

\clearpage

\clearpage

\clearpage

\clearpage

\clearpage

\clearpage

\clearpage
\begin{figure}
\includegraphics[scale=0.9]{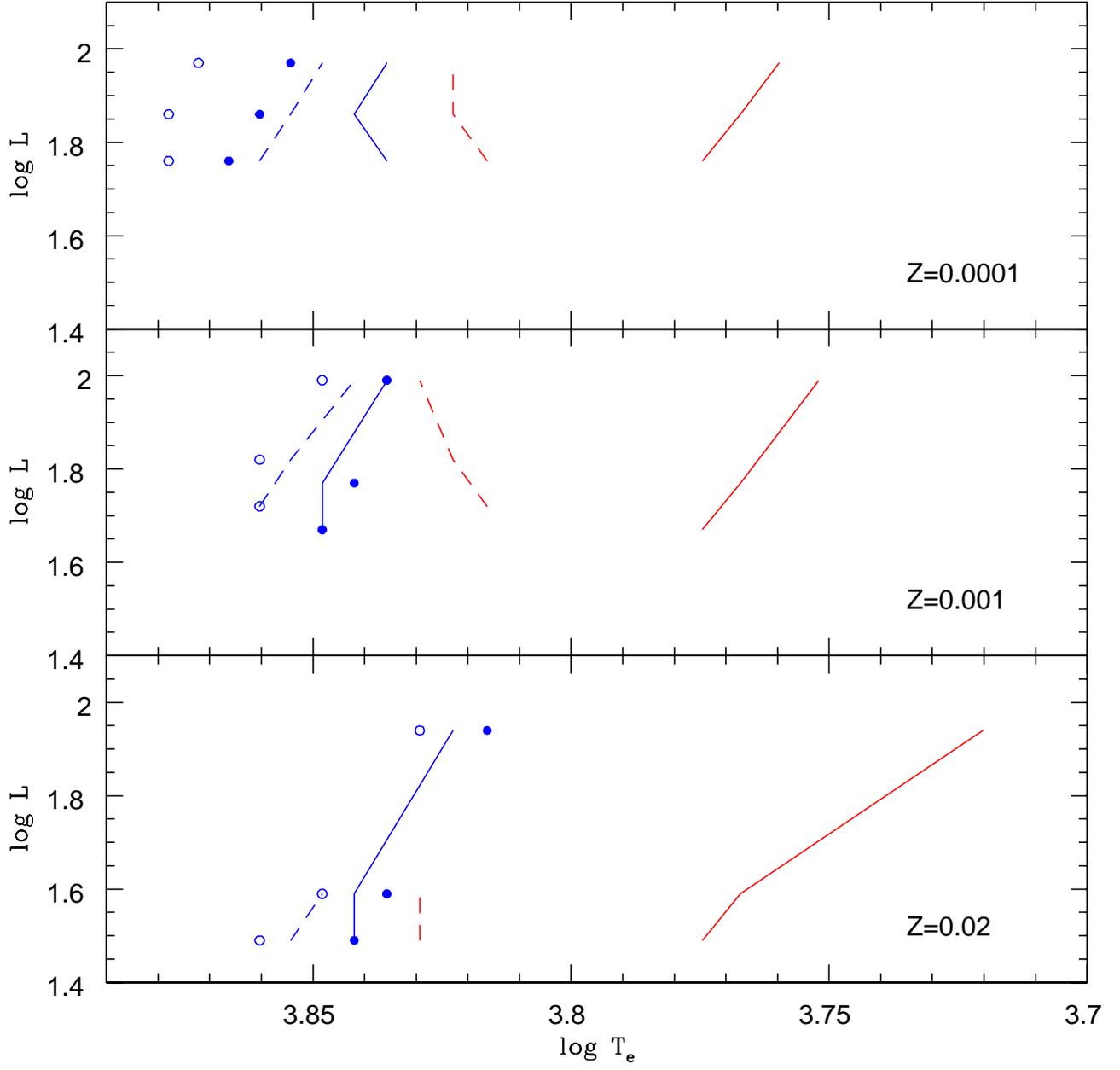}
\caption{From top to bottom the panels display predicted instability strips for 
FU (solid) and FO (dashed) pulsators at three different chemical compositions: 
$Z=0.0001$ (top panel), $Z=0.001$ (middle) and $Z=0.02$ (bottom). The hot edges 
are plotted in blue, while the cool edges in red.
The linear FOBE and FBE are represented by open and filled blue
circles, respectively (see Appendix).} \label{strip}
\end{figure}

\clearpage
\begin{figure}
\includegraphics[scale=0.9]{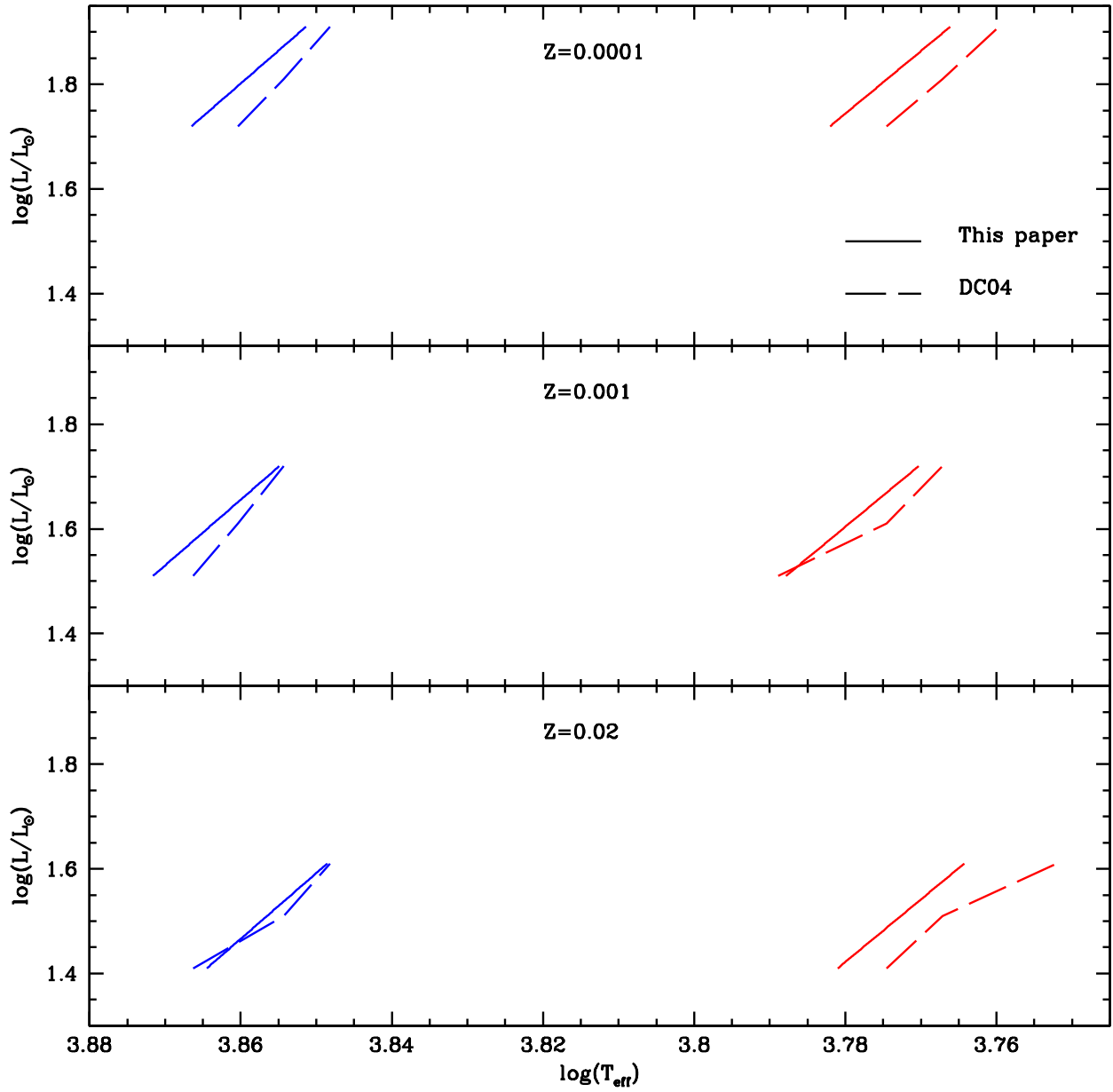}
\caption{Comparison between the current instability strip boundaries 
(solid lines) and similar predictions by~\citet{dmc04} (dashed lines).
The blue lines display the FO blue boundaries, while the red lines the FU red 
boundaries.}\label{strip_conf}
\end{figure}

\clearpage
\begin{figure}
\includegraphics[scale=0.9]{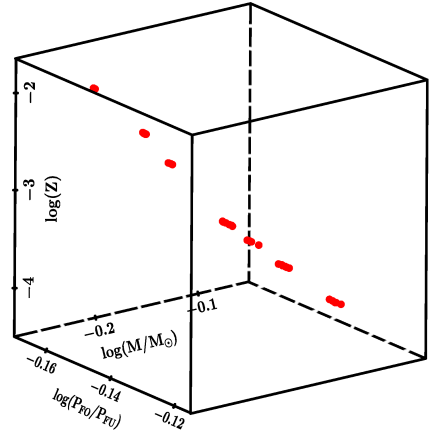}
\caption{Tridimensional plot showing the correlation among stellar mass 
($\log (M/M_{\odot})$), metallicity ($\log (Z)$) and period ratios 
($\log (P_{FO}/P_{FU})$) for the pulsation models located inside the 
so-called "OR region".} \label{fig_rel_mass}
\end{figure}

\clearpage
\begin{figure}
\includegraphics[scale=0.9]{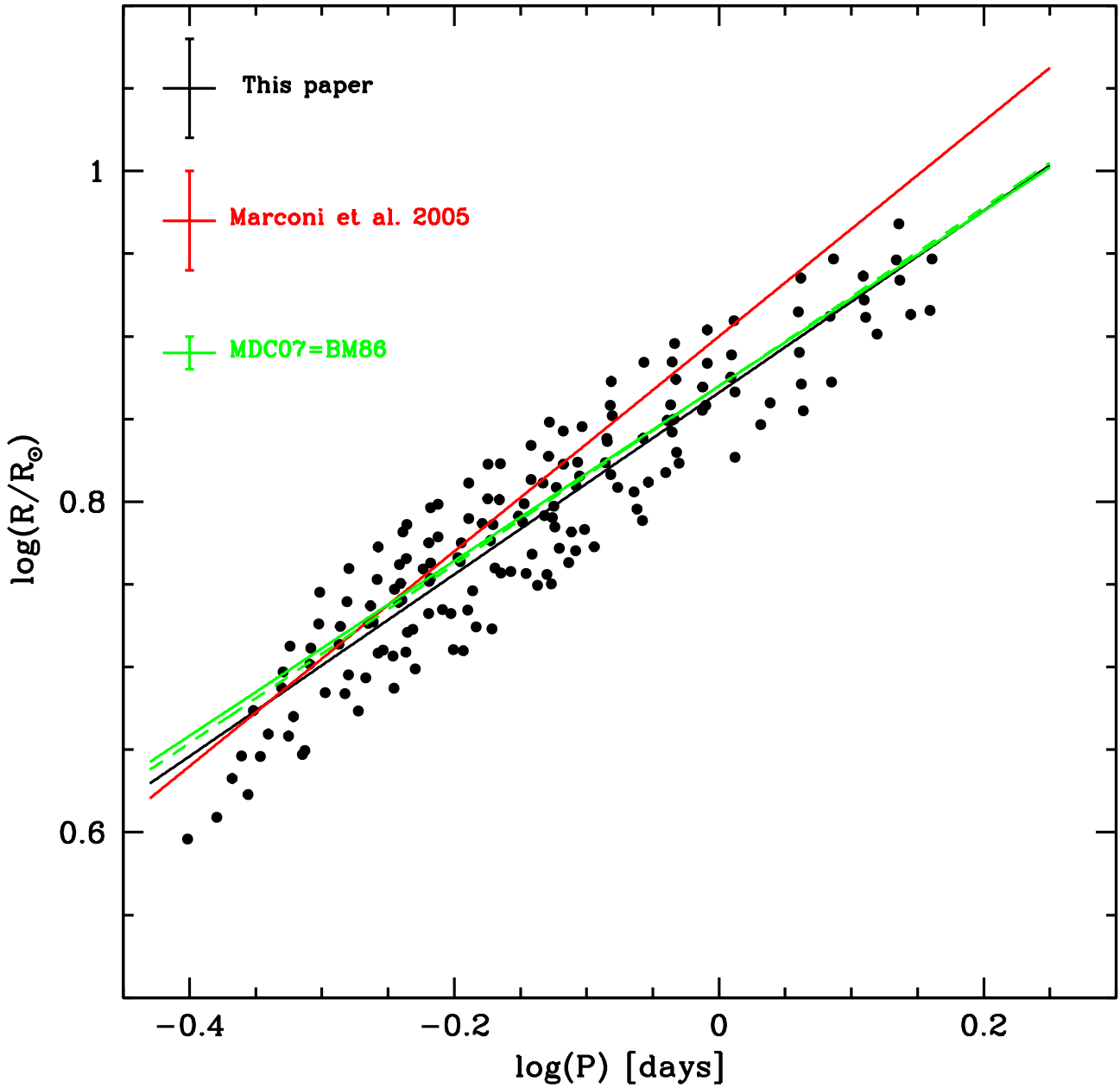}
\caption{Period-Radius relation for FU pulsators covering the entire range of 
chemical compositions adopted in the current investigation. The black solid 
line depicts the linear regression with a standard deviation of 0.03 dex. 
The red solid line shows the theoretical PR relation provided 
by~\citet{marconi05}, while the solid green line the extrapolation to shorter 
periods of the PR for BL Herculis variables based on pulsation models provided 
by~\citet{m07}. The latter is almost identical to the empirical PR relation provided 
by~\citet{bm86} for the same class of variable stars and represented
in the plot by the dashed green line.} \label{pr_f}
\end{figure}

\clearpage
\begin{figure}
\includegraphics[scale=0.9]{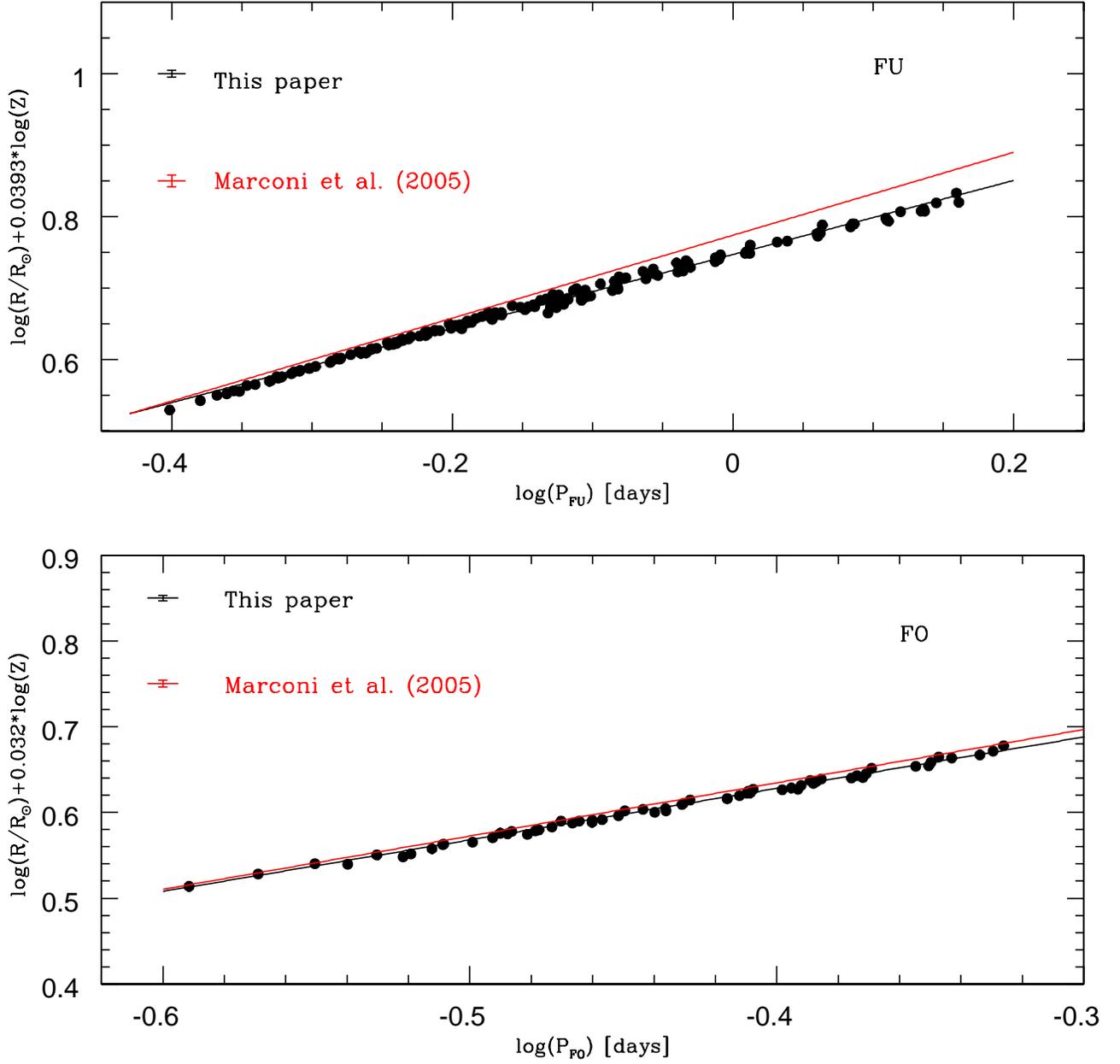}
\caption{Period-Radius-Metallicity (PRZ) relations for FU (top panel) and
FO (bottom panel) pulsators with a standard deviation of 0.006 and 0.003 dex, 
respectively. Black lines display the linear regression based on the 
current models, while the red lines show similar PRZ relations provided 
by~\citet{marconi05}.}\label{pr_f_fo}
\end{figure}

\clearpage
\begin{figure}
\includegraphics[scale=0.8]{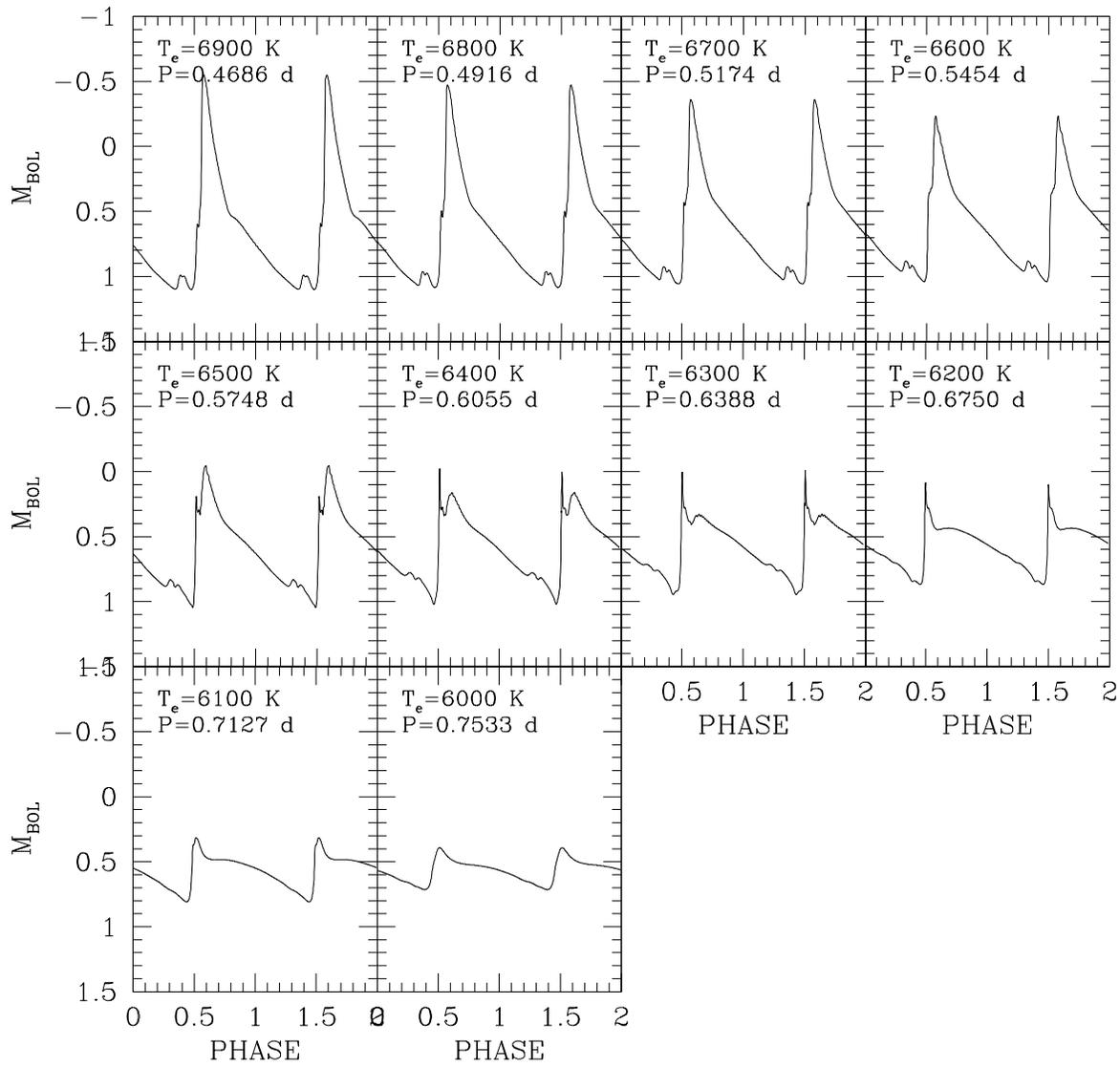}
\caption{Predicted bolometric light-curves for FU pulsators at fixed chemical 
composition ($Z=0.0006$, $Y=0.245$). The models refer to sequence
($M=0.67M_{\odot}$, $\log{L/L_{\odot}}=1.69$). The effective temperatures and 
the pulsation periods are labeled.}\label{fig_cdl1}
\end{figure}

\begin{figure}
\includegraphics[scale=0.8]{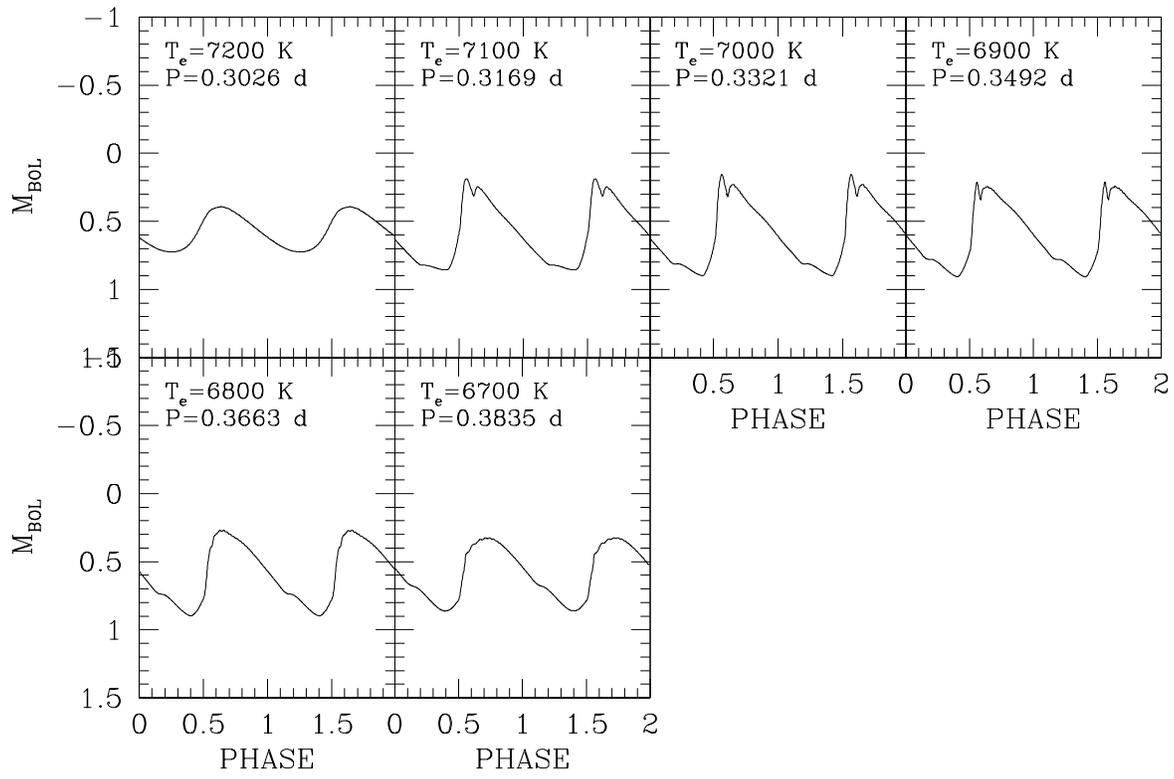}
\caption{Same as Figure~\ref{fig_cdl1}, but for FO pulsators.}\label{fig_cdl2}
\end{figure}

\clearpage
\begin{figure}
\includegraphics[scale=0.9]{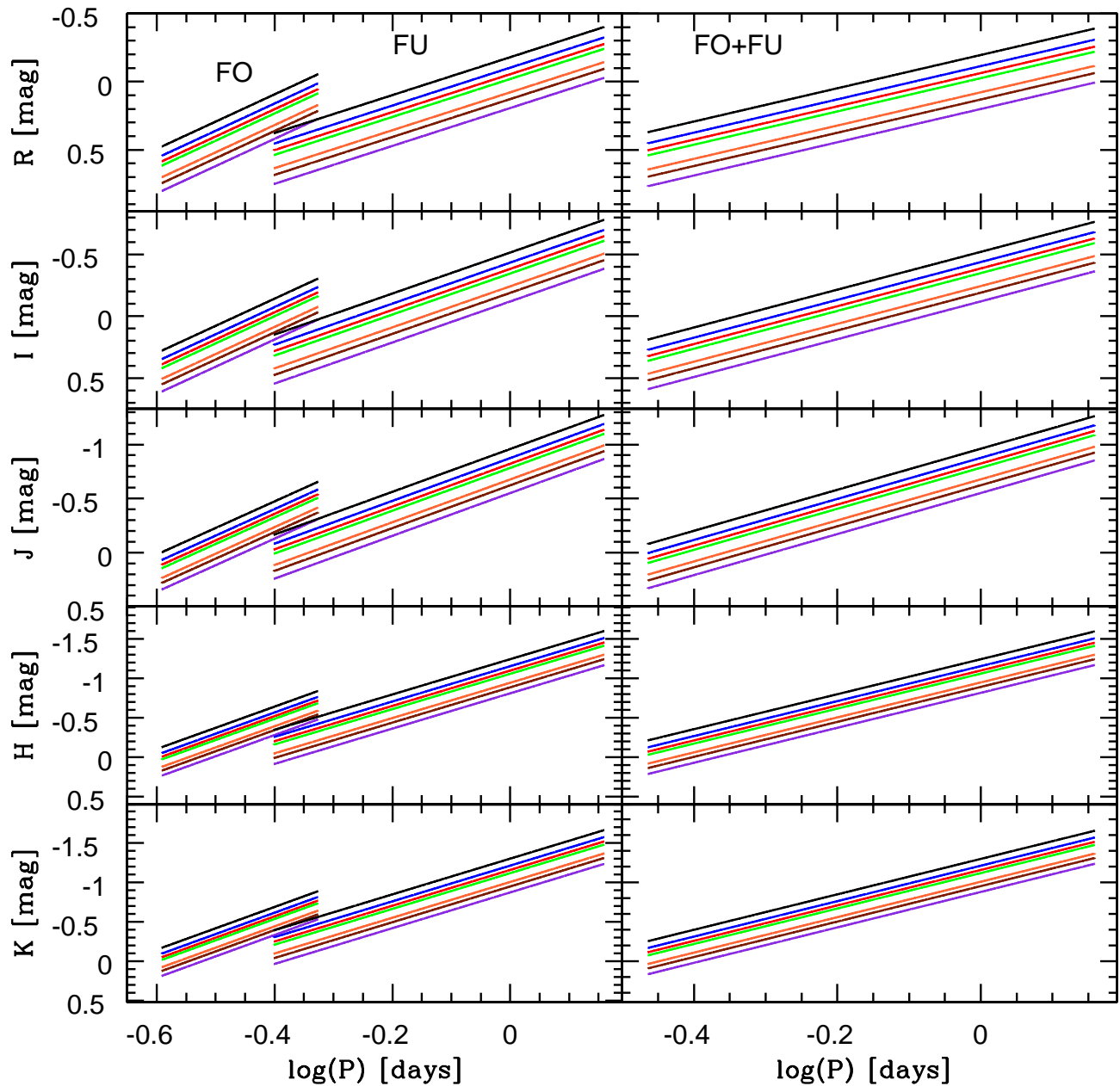}
\caption{Left panels -- Predicted metal-dependent optical and NIR ($RIJHK$) 
PL relations FU and FO pulsators. Lines of different colors display predictions 
with metal abundances ranging from $Z=0.0001$ (black lines) to $Z=0.02$ 
(purple lines).
Right panels -- Same as the left ones, but for FU$+$FO pulsators.}\label{pl1}
\end{figure}

\clearpage
\begin{figure}
\includegraphics[scale=0.9]{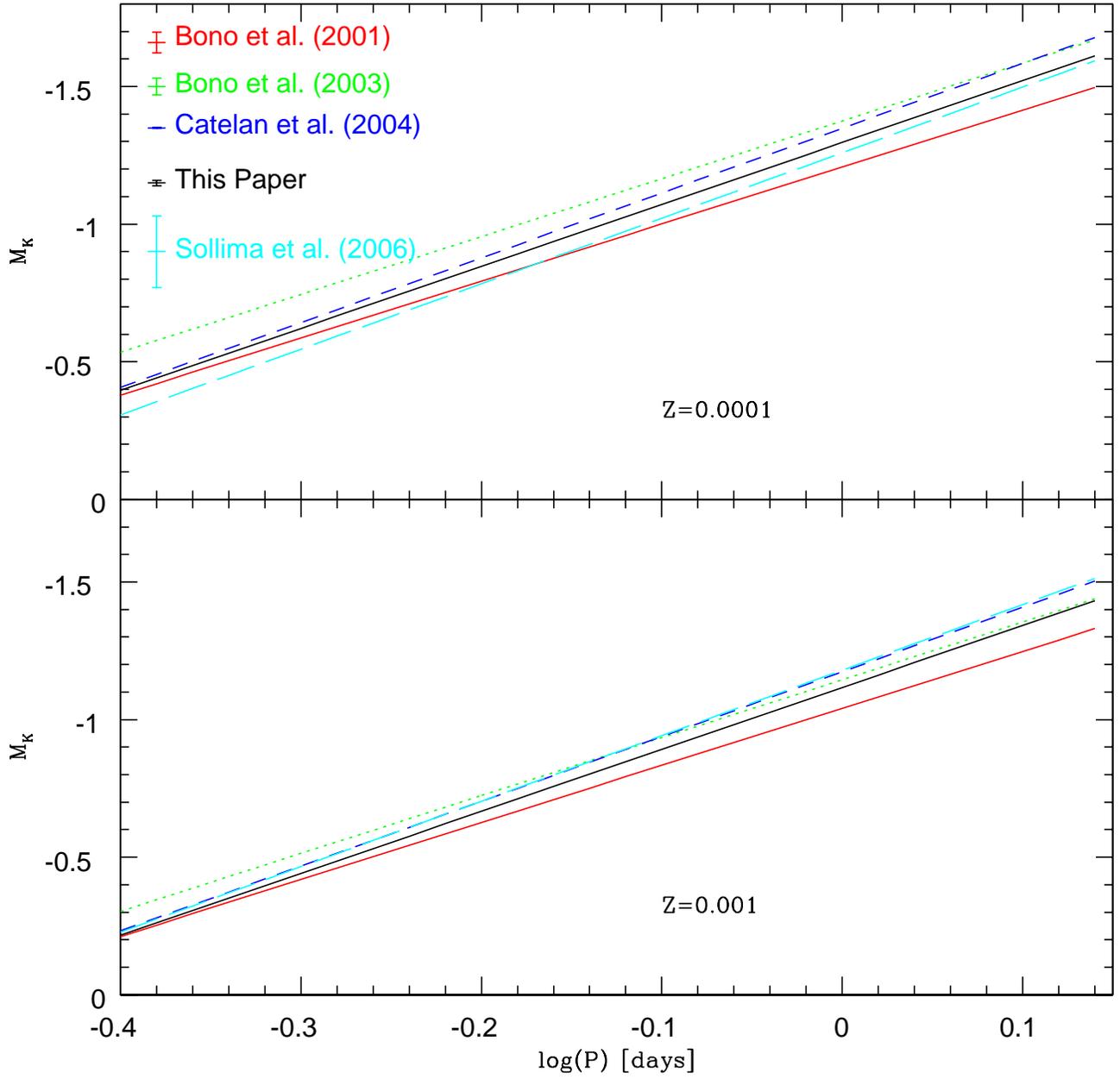}
\caption{Comparison among the current PLZ relations for the two different 
metal abundances (Z=0.0001, top; Z=0.001, bottom) and similar relations 
available in the literature. The standard deviations of the above PLZ relations, 
when available, are plotted in the top left corner of the top panel.} \label{pl_conf}
\end{figure}

\clearpage
\begin{figure}
\includegraphics[scale=0.9]{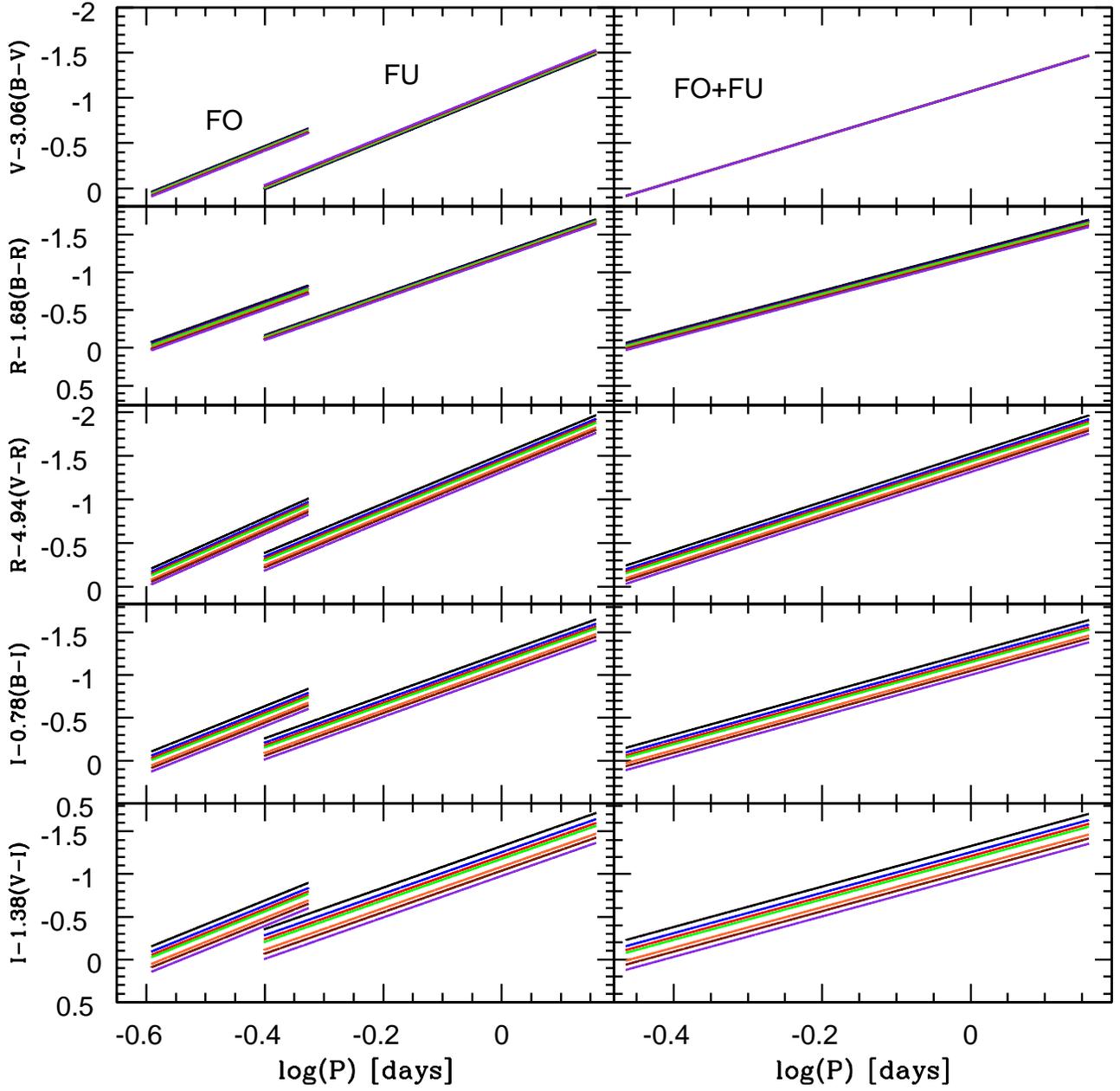}
\caption{Left panels -- Predicted metal-dependent optical PWZ relations. The color 
coding is the same as in Figure~\ref{pl1}. The coefficients of the color terms 
are labeled on the Y-axis. Note either the minimal or the marginal metal dependence 
of the PW(V,B-V) (top panel) and of the PW(R,B-R) (second panel from top).
Right panels -- Same as the left ones, but for FU$+$FO pulsators.}\label{pw}
\end{figure}

\clearpage
 \begin{figure}
 \includegraphics[scale=0.9]{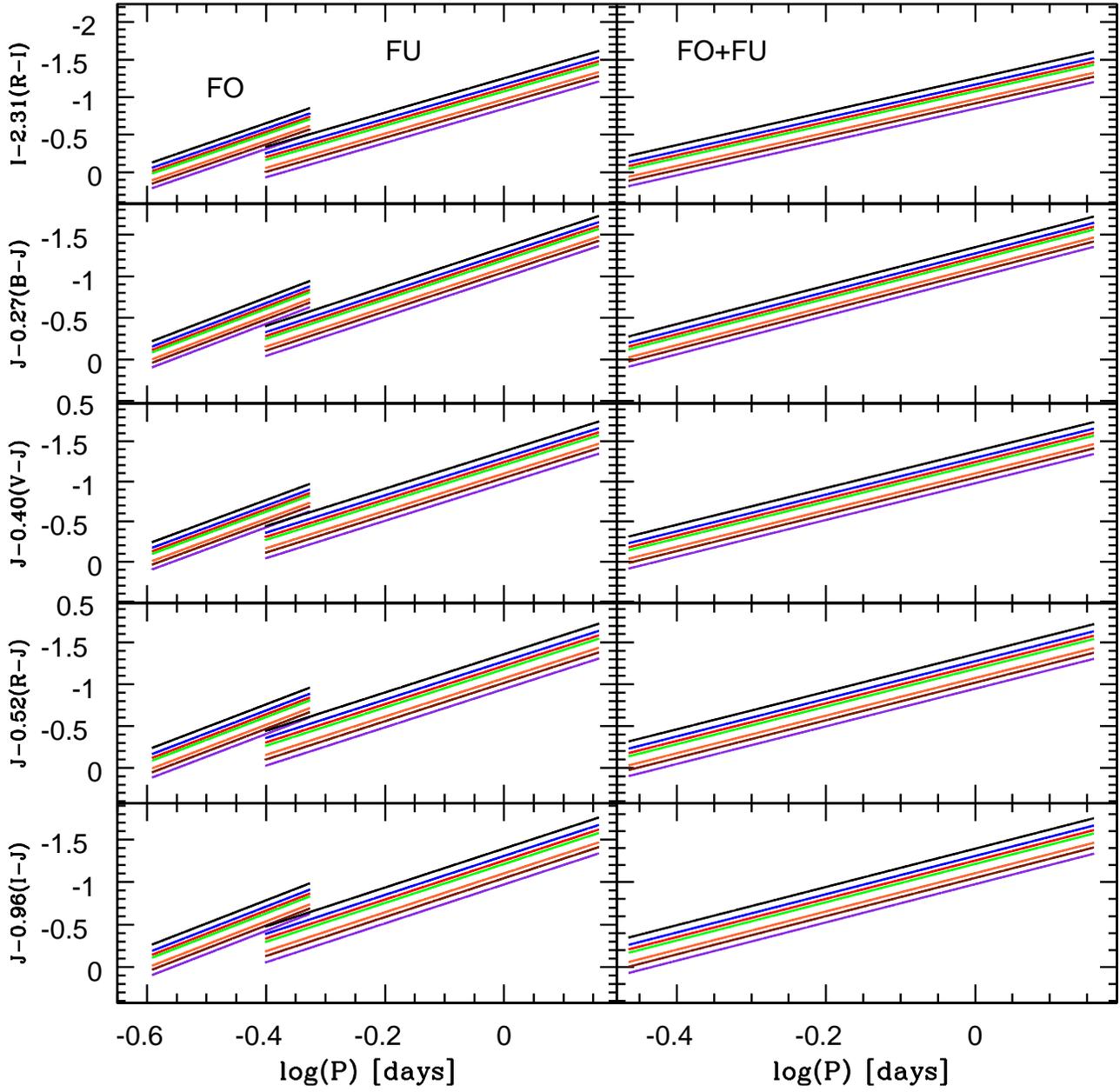} 
 \caption{Same as in Figure~\ref{pw}, but for optical-NIR PWZ relations.}\label{pw1}
 \end{figure}

\clearpage
\begin{figure}
 \includegraphics[scale=0.9]{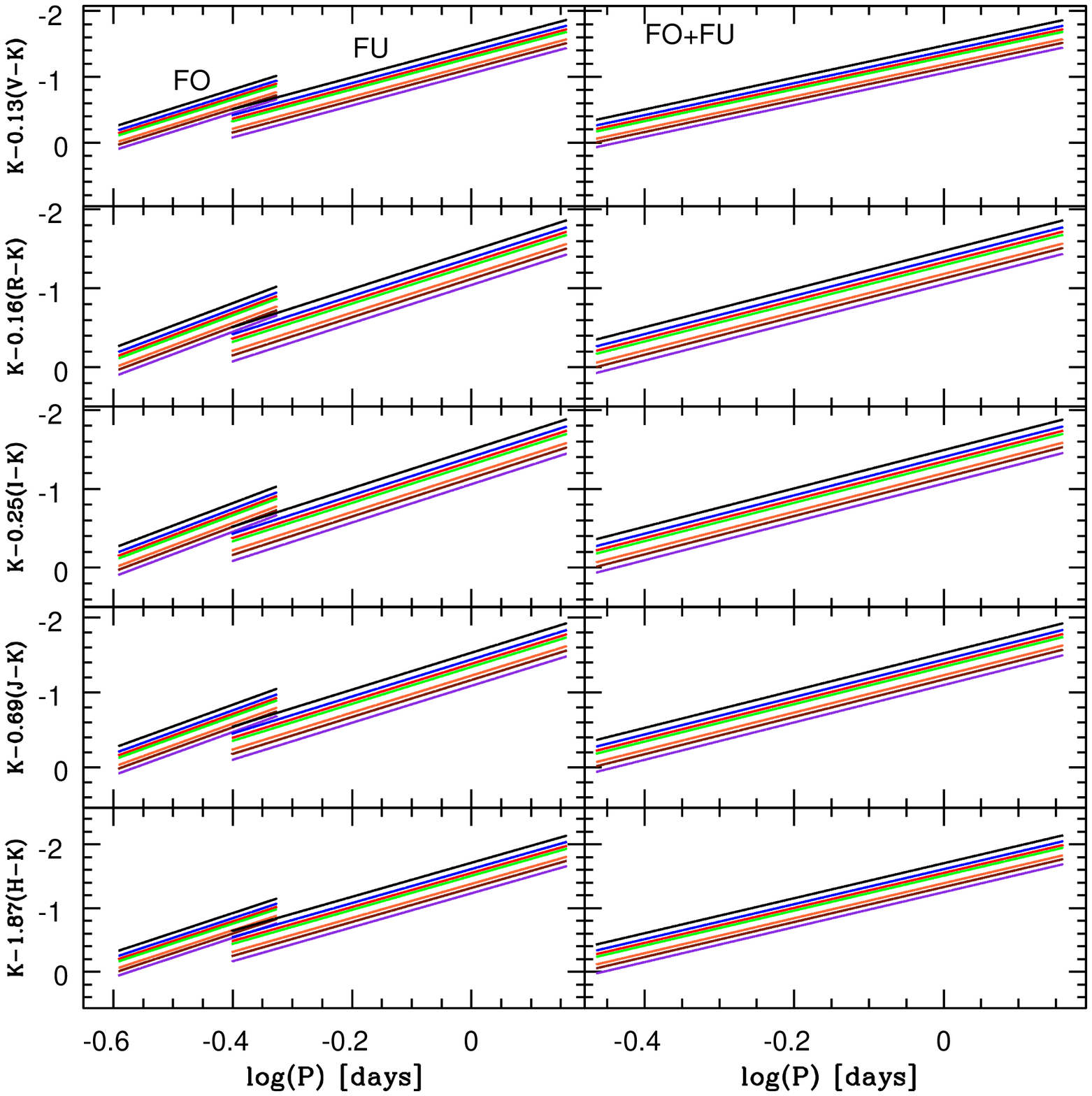} 
 \caption{Same as in Figure~\ref{pw}, but for NIR PWZ relations.}\label{pw2}
 \end{figure}

\clearpage
\begin{figure}
\includegraphics[scale=0.9]{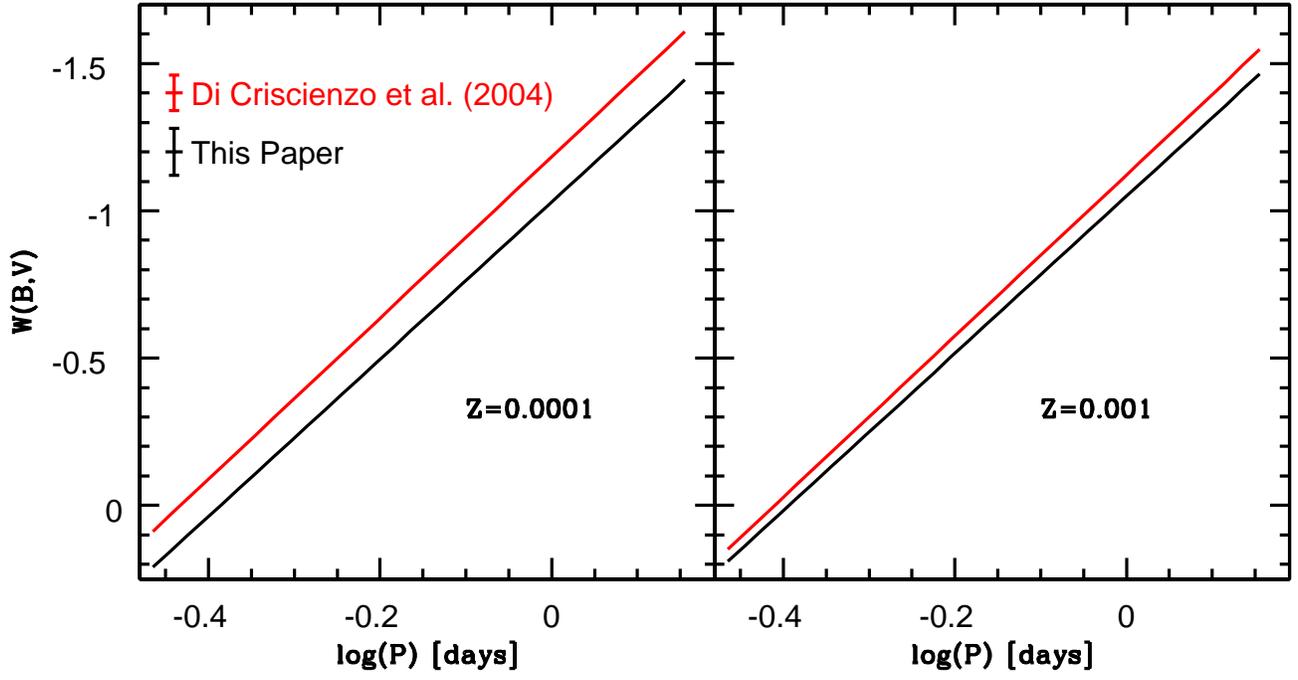}
\caption{Comparison between the current PWZ(V,B-V) relations and 
similar relations provided by~\citet{dmc04} for two different 
metal abundances: $Z=0.0001$ (top panel) and $Z=0.001$ 
(bottom panel).} \label{plw_conf}
\end{figure}

\clearpage
\begin{figure}
\includegraphics[scale=0.9]{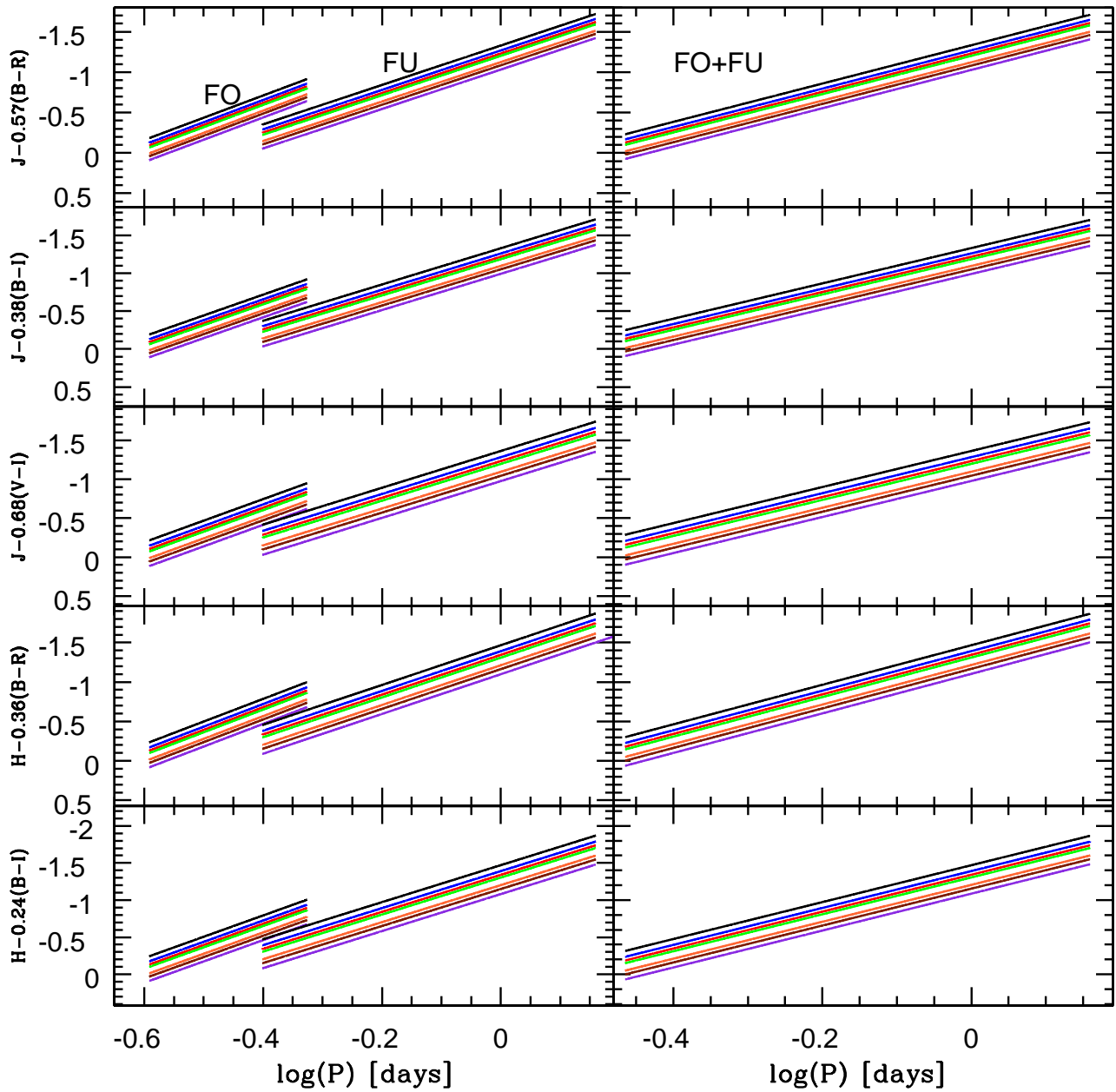}
\caption{Left Panels -- Predicted metal-dependent triple bands optical-NIR 
PWZ relations.  The color coding is the same as in Figure~\ref{pw}. Right panels -- 
Same as the left ones, but for FU$+$FO pulsators.} \label{plw_3bands}
\end{figure}

\clearpage

\clearpage

\begin{appendix}

\section{Linear models}

 The current set of linear models are purely radiative \citep{bs94} and do not provide an estimate of the red boundaries 
of the instability strip. The physical mechanism that is providing the 
quenching of radial oscillation is the increased efficiency of 
convective transport. However they provide the envelope structures and 
the linear radial eigenfunctions adopted by the nonlinear hydrodynamical models. 
The nonlinear analysis is performed by imposing a constant perturbation 
velocity (5-10 km sec$^{-1}$) both to the FU and the FO linear radial 
eigenfunctions \citep{bs94,bms99}. 
After this initial perturbation the dynamical behavior is followed in time 
until the radial displacements approach their asymptotic behavior. 
In the current linear models the inner boundary of the static model is 
located at a distance of $\approx$10\% of the equilibrium photospheric 
radius. The outer boundary is located at an optical depth $\tau=0.0001$ and 
the stellar mass attached to the outermost zone is at least one order of 
magnitude larger than the mass of the surface. The mass zoning of the 
stellar envelope was fixed assuming a mass ratio between consecutive zones 
of 1.04 for temperatures cooler than 60,000 K. It was increased by 0.001 in 
the zones located at higher temperatures. As a whole we ended up 
with envelope models including typically 200-300 discrete mesh points \citep{b98}, allowing smooth variations of the physical 
parameters in both driving and damping envelope regions. The envelope mass 
decreases from 18\% to a few percents of the total mass when moving from the 
blue faint to the red bright pulsation models. The dependence of the envelope 
mass on chemical composition is negligible.

Fig.~\ref{strip} shows the comparison between the nonlinear edges of the instability 
strip (\S~4) and the linear FU (filled circles) and FO (open circles) blue boundaries. We note that the 
linear FOBE and FBE shows a more evident dependence on the metal content 
than their nonlinear counterpart when moving from Z=0.0001 to Z=0.02. 
In particular, for the most metal-poor chemical 
composition (Z=0.0001, Y=0.245) the linear boundaries are significantly 
bluer than the nonlinear ones. This is an expected result, because the 
inclusion of convective transport causes a decrease in pulsation 
destabilization (driving), therefore, the blue edges move towards
lower effective temperatures. The quenching caused by convection, close 
to the blue edges of the instability strip, becomes less efficient in the 
metal-rich regime. It is partially counterbalanced by the  
driving of the K-bump, i.e. the radiative opacity peak located at 
$\approx$250,000 K \citep{sea94,ri92,bim96}.

\end{appendix}

\end{document}


\scriptsize
\begin{deluxetable}{lllllll}
\tabletypesize{\scriptsize}
\tablewidth{0pt}
\tablecaption{Evolutionary lifetimes for typical ZAHB and MSTO 
stellar structures at different metal abundances.\label{table_time}}
 \tablehead{
 \multicolumn{1}{c}{Z\tablenotemark{a}} &
 \multicolumn{1}{c}{Y\tablenotemark{a}} &
 \multicolumn{1}{c}{T$_{MS}$\tablenotemark{b}} &
 \multicolumn{1}{c}{T$_{HB}$\tablenotemark{c}} &
 \multicolumn{1}{c}{T$_{HB}/$T$_{MS}$} &
 \multicolumn{1}{c}{M$_{HB}$\tablenotemark{d}} &
 \multicolumn{1}{c}{M$_{TO}$\tablenotemark{e}} 
 }
 \startdata
 0.0001 & 0.245 & 2056.59 & 66.80  & 0.03248  & 0.80 & 0.80\\
 0.0010 & 0.246 & 3433.67 & 77.545 & 0.02258  & 0.65 & 0.79\\
 0.0080 & 0.256 & 5969.38 & 86.101 & 0.01442  & 0.57 & 0.85
 \enddata
\tablenotetext{a}{Metal (Z) and helium (Y) abundance by mass.}
\tablenotetext{b}{Central hydrogen burning lifetime between a point that 
is 0.25 dex fainter than the MSTO and the MSTO (Myr).}
\tablenotetext{c}{Central helium burning lifetime at 90\% of core helium 
exhaustion (Myr).}
\tablenotetext{d}{ZAHB stellar mass roughly located at the center of the 
instability strip (solar units).}
\tablenotetext{e}{Stellar mass at the MSTO for an of 13 Gyrs (solar units).}
\end{deluxetable}


\begin{deluxetable}{ccccccc}
\tabletypesize{\scriptsize}
\tablewidth{0pt}
\tablecaption{Intrinsic stellar parameters adopted to constract the pulsation 
models and predicted effective temperatures of the instability strip boundaries. 
The different sets of pulsation models were computed assuming the ZAHB mass 
value (column 1) and three different luminosity levels (column 2): the 
ZAHB level (sequence A), a luminosity level 0.1 dex brighter than the 
ZAHB (sequence B) and the luminosity level of central He exhaustion 
(sequence D). An additional sequence of RRL models was computed assuming 
a stellar mass $\sim$10\% smaller than the ZAHB mass value and 0.2 dex 
brighter than the ZAHB luminosity level (sequence C).\label{PARAMETRI}}
 \tablehead{
 \multicolumn{1}{c}{$M_{ZAHB}$\tablenotemark{a}} &
 \multicolumn{1}{c}{$\log L/L_{\odot}$\tablenotemark{b}} &
\multicolumn{1}{c}{$Sequence$\tablenotemark{c}} &
 \multicolumn{1}{c}{$FOBE$\tablenotemark{d}} &
 \multicolumn{1}{c}{$FBE$\tablenotemark{d}} &
 \multicolumn{1}{c}{$FORE$\tablenotemark{d}} &
 \multicolumn{1}{c}{$FRE$\tablenotemark{d}} 
 }
\startdata
\multicolumn{7}{c}{\textbf{Z=0.0001 Y=0.245}} \\
  0.80 & 1.76 & A & 7250 &6850  & 6550 & 5950\\
  0.80 & 1.86 & B & 7150& 6950 &6650 &5850 \\
  0.80 & 1.97 & D &7050& 6850 &6650 &5750 \\
  0.72 & 1.96 & C & 6950  & 6850 & 6650 & 5750 \\  
\multicolumn{7}{c}{\textbf{Z=0.0003 Y=0.245}} \\
0.716 & 1.72 & A &7250 &6950 &6550 &5950\\
0.716 & 1.82 & B &7150& 6950& 6650& 5850\\
\enddata
 \tablenotetext{a}{ZAHB stellar mass (solar units).}
 \tablenotetext{b}{Logarithmic luminosity (solar units).}
 \tablenotetext{c}{Name of the sequence of pulsation models 
(see Figure~\ref{fig_cmd} for more details).}
 \tablenotetext{d}{Effective temperature (K) of the instability strip 
boundaries: FOBE, FO blue edge; FBE, FU blue edge; FORE, FO red edge; 
FRE, FU red edge.}
 \end{deluxetable}

\begin{deluxetable}{cccccccccccc}
\tabletypesize{\scriptsize}
\tablewidth{0pt}
\tablecaption{
Intensity-averaged mean magnitudes for the entire grid of FU models. 
For each set of models we list the chemical composition (metal and helium 
content), the stellar mass and the luminosity levels adopted for the different 
sequences. The first three columns give, for every pulsation model, the 
effective temperature, the logarithmic luminosity level and the logarithmic 
period. The last eight columns list the predicted optical Johnson-Cousins 
$U,B,V,R,I$ and the NIR 2MASS $J,H,K$ intensity-averaged mean 
magnitudes. The last column gives the mean radius in solar units.\label{medieintF}}
\tablehead{
\multicolumn{1}{c}{\textbf{T$_{e}$}} &
\multicolumn{1}{c}{\textbf{$\log L$}} &
\multicolumn{1}{c}{\textbf{$\log P$}} &
\multicolumn{1}{c}{\textbf{U}} & \multicolumn{1}{c}{\textbf{B}} &
\multicolumn{1}{c}{\textbf{V}} & \multicolumn{1}{c}{\textbf{R}} &
\multicolumn{1}{c}{\textbf{I}} & \multicolumn{1}{c}{\textbf{J}} &\multicolumn{1}{c}{\textbf{H}} &
\multicolumn{1}{c}{\textbf{K}} & \multicolumn{1}{c}{\textbf{$\log (R/R_{\odot})$}}  
} 
\startdata

 \multicolumn{12}{c}{\textbf{Z=0.0001 Y=0.245 M=0.80 M$_{\odot}$}} \\

 6800 &  1.76 & -0.3016 &  0.653 &  0.652 &  0.410 &  0.225&  -0.009 & -0.296& -0.540& -0.559& 0.7452\\
 6700 &  1.76 & -0.2795 &  0.672 &  0.670 &  0.407 &  0.210&  -0.035 & -0.337& -0.592& -0.612& 0.7595\\
 6600 &  1.76 & -0.2574 &  0.691 &  0.695 &  0.409 &  0.198&  -0.061 & -0.376& -0.644& -0.665& 0.7726\\
 6500 &  1.76 & -0.2358 &  0.710 &  0.727 &  0.416 &  0.189&  -0.084 & -0.414& -0.696& -0.718& 0.7861\\
 6400 &  1.76 & -0.2124 &  0.731 &  0.763 &  0.426 &  0.184&  -0.104 & -0.450& -0.746& -0.769& 0.7984\\
 6300 &  1.76 & -0.1892 &  0.750 &  0.797 &  0.438 &  0.180&  -0.122 & -0.483& -0.793& -0.817& 0.8112\\
 6200 &  1.76 & -0.1650 &  0.768 &  0.826 &  0.447 &  0.177&  -0.137 & -0.513& -0.835& -0.861& 0.8229\\
\enddata
\end{deluxetable}

\begin{deluxetable}{cccccccccccc}
\tabletypesize{\scriptsize}
\tablewidth{0pt}
\tablecaption{Same as Table~\ref{medieintF}, but for FO models.\label{medieintFO}} 
\tablehead{
\multicolumn{1}{c}{\textbf{T$_{e}$}} &
\multicolumn{1}{c}{\textbf{$\log L$}} &
\multicolumn{1}{c}{\textbf{$\log P$}} &
\multicolumn{1}{c}{\textbf{U}} & \multicolumn{1}{c}{\textbf{B}} &
\multicolumn{1}{c}{\textbf{V}} & \multicolumn{1}{c}{\textbf{R}} &
\multicolumn{1}{c}{\textbf{I}}  &\multicolumn{1}{c}{\textbf{J}} &
\multicolumn{1}{c}{\textbf{H}} &
\multicolumn{1}{c}{\textbf{K}} & \multicolumn{1}{c}{\textbf{$log (R/R_{\odot})$}} 
}
\startdata

  \multicolumn{12}{c}{\textbf{Z=0.0001 Y=0.245 M=0.72 M$_{\odot}$}} \\

 7200  & 1.76 & -0.5088  & 0.613 &  0.562  & 0.366 &  0.223 &  0.040 & -0.182 &-0.351  & -0.367 &0.6907\\
 7100  & 1.76 & -0.4901  & 0.620 &  0.583  & 0.378 &  0.226 &  0.030 & -0.209 &-0.398 	& -0.443 &0.7039\\
 7000  & 1.76 & -0.4703  & 0.632 &  0.604  & 0.382 &  0.218 &  0.010 & -0.243 &-0.446 	& -0.463 &0.7178\\
 6900  & 1.76 & -0.4495  & 0.645 &  0.625  & 0.385 &  0.209 & -0.012 & -0.279 &-0.494	& -0.512 &0.7299\\
 6800  & 1.76 & -0.4282  & 0.660 &  0.652  & 0.389 &  0.198 & -0.036 & -0.316 &-0.544 	& -0.563 &0.726\\
 6700  & 1.76 & -0.4077  & 0.675 &  0.682  & 0.396 &  0.191 & -0.058 & -0.352 &-0.592 	& -0.612 &0.7548\\
 6600  & 1.76 & -0.3855  & 0.689 &  0.711  & 0.405 &  0.186 & -0.076 & -0.385 &-0.638 	& -0.659 &0.7670\\
\enddata
\end{deluxetable}

\begin{deluxetable}{llllll}
\tabletypesize{\scriptsize}
\tablewidth{0pt}
\tablecaption{Analytical relations for the edges of the instability strip 
into the observational plane. They have the form: 
Color$_{edge}$=$a$*Mag + $b$*$\log$ Z + $c$.\label{bordicol}}
\tablehead{
 \multicolumn{1}{c}{Edge} &
 \multicolumn{1}{c}{Mag} &
 \multicolumn{1}{c}{Color} &
 \multicolumn{1}{c}{a} &
 \multicolumn{1}{c}{b} &
 \multicolumn{1}{c}{c}
} 
 \startdata
 FOBE & M$_V$  & $(B-V)$ & -0.05$\pm$0.02 & 0.05$\pm$0.01 & 0.39$\pm$0.01 \\
      & \ldots & $(V-I)$ & -0.09$\pm$0.01 & 0.03$\pm$0.01 & 0.45$\pm$0.01 \\
      & M$_K$  & $(J-K)$ & -0.03$\pm$0.01 & \ldots & 0.17$\pm$0.01 \\
 FRE  & M$_V$  & $(B-V)$ & -0.12$\pm$0.07 & 0.15$\pm$0.03 & 0.98$\pm$0.11 \\
      & \ldots & $(V-I)$ & -0.13$\pm$0.02 & 0.04$\pm$0.01 & 0.86$\pm$0.03 \\
      & M$_K$  & $(J-K)$ & -0.07$\pm$0.01 & \ldots & 0.32$\pm$0.01
 \enddata
 \end{deluxetable}

\begin{deluxetable}{lcccccccc}
\tabletypesize{\scriptsize}
\tablewidth{0pt}
\tablecaption{Coefficients of the predicted metal-dependent optical and
NIR ($RIJHK$) PLZ relations for FU, FO and FU$+$FO pulsators. They take 
account of the entire metallicity range (Z=0.0001--0.02). The global 
(FU$+$FO) relations were derived by fundamentalizing the FO periods 
using the canonical relation.\label{tab_plz1}}
 \tablehead{
 \multicolumn{1}{c}{Type\tablenotemark{a}} &
 \multicolumn{1}{c}{Band\tablenotemark{b}} &
 \multicolumn{1}{c}{$a$\tablenotemark{c}} &
 \multicolumn{1}{c}{$b$\tablenotemark{d}} &
 \multicolumn{1}{c}{$c$\tablenotemark{e}} &
 \multicolumn{1}{c}{$\sigma_{a}$\tablenotemark{f}} &
 \multicolumn{1}{c}{$\sigma_{b}$\tablenotemark{f}} &
 \multicolumn{1}{c}{$\sigma_{c}$\tablenotemark{f}} &
 \multicolumn{1}{c}{$rms$\tablenotemark{g}} }
 \startdata
     FU  &R &   0.24 &  -1.39 & 0.17   &   0.02 & 0.07 &  0.01 &   0.10\\
     FO  &R &  -0.34 &  -2.00 & 0.14   &   0.07 & 0.10 &  0.01 &   0.07\\
     Glob&R &   0.25 &  -1.22 & 0.17   &   0.02 & 0.06 &  0.01 &   0.10\\  
     FU  &I &  -0.07 &  -1.66 & 0.17   &   0.02 & 0.06 &  0.01 &   0.09\\
     FO  &I &  -0.65 &  -2.20 & 0.14   &   0.06 & 0.10 &  0.01 &   0.05\\
     Glob&I &  -0.07 &  -1.53 & 0.17   &   0.02 & 0.04 &  0.01 &   0.09\\     
     FU  &J &  -0.51 &  -1.98 & 0.17   &   0.01 & 0.04 &  0.01 &   0.06\\
     FO  &J &  -1.07 &  -2.46 & 0.15   &   0.04 & 0.08 &  0.01 &   0.04\\
\enddata
 \tablenotetext{a}{Pulsation mode.}
 \tablenotetext{b}{Photometric band.}
 \tablenotetext{c}{Zero-point of the PLZ relation.}
 \tablenotetext{d}{Coefficient of the logarithmic period.}
 \tablenotetext{e}{Coefficient of the metallicity term.}
 \tablenotetext{e}{Errors on coefficients.}
 \tablenotetext{f}{Standard deviation.}
 \end{deluxetable}


\begin{deluxetable}{llcccccccc}
\tabletypesize{\scriptsize}
\tablewidth{0pt}
\tablecaption{{\bf Coefficients of the predicted dual-band, metal-dependent, optical and NIR PWZ
relations for FU, FO and FU+FO pulsators. They take account of the entire metallicity range
(Z=0.0001–0.02). The global (FU+FO) relations were derived by fundamentalizing the FO
periods using the canonical relation.} \label{tab_pwz}
}
\tablehead{
\multicolumn{1}{c}{Mode\tablenotemark{a}} &
\multicolumn{1}{c}{Band\tablenotemark{b}} &
\multicolumn{1}{c}{$\xi$\tablenotemark{c}}&
\multicolumn{1}{c}{$a$\tablenotemark{d}}  &
\multicolumn{1}{c}{$b$\tablenotemark{e}}  & 
\multicolumn{1}{c}{$c$\tablenotemark{f}}  &
\multicolumn{1}{c}{$\sigma_{a}$\tablenotemark{g}} &
\multicolumn{1}{c}{$\sigma_{b}$\tablenotemark{g}} &
\multicolumn{1}{c}{$\sigma_{c}$\tablenotemark{g}} &
\multicolumn{1}{c}{$rms$\tablenotemark{h}}
}
\startdata
 FU  &V,B-V & 3.06 &  -1.11 & -2.67 & -0.02 &   0.02  & 0.04 &  0.01 &  0.07\\
 FO  &V,B-V & 3.06 &  -1.47 & -2.65 &  0.02 &   0.05  & 0.09 &  0.01 &  0.04\\
 Glob&V,B-V & 3.06 &  -1.07 & -2.49 &  0.01 &   0.01  & 0.04 &  0.01 &  0.08\\
 FU  &R,B-R & 1.68 &  -1.19 & -2.74 &  0.03 &   0.01  & 0.04 &  0.01 &  0.07\\
 FO  &R,B-R & 1.68 &  -1.62 & -2.82 &  0.05 &   0.04  & 0.08 &  0.01 &  0.04\\
 Glob&R,B-R & 1.68 &  -1.17 & -2.61 &  0.04 &   0.01  & 0.03 &  0.01 &  0.07\\
 FU  &R,V-R & 4.94 &  -1.30 & -2.82 &  0.09 &   0.01  & 0.04 &  0.01 &  0.06\\
 FO  &R,V-R & 4.94 &  -1.79 & -3.02 &  0.08 &   0.03  & 0.07 &  0.01 &  0.03\\
\enddata
 \tablenotetext{a}{Pulsation mode.}
 \tablenotetext{b}{Photometric band.}
 \tablenotetext{c}{{\bf Coefficient of the color term. They have been
   estimated using the \citet{cardelli}  reddining law and the central
   wavelengths of the Landolt optical photometric system \citep{landolt} and of the 2MASS NIR photometric system \citep{cutri}. The above changes also imply a total to selective absorption ratio $R_V=3.06$. The new assumptions will be addressed in a forthcoming paper.}}
 \tablenotetext{d}{Zero-point of the PLZ relation.}
 \tablenotetext{e}{Coefficient of the logarithmic period.}
 \tablenotetext{f}{Coefficient of the metallicity term.}
 \tablenotetext{g}{Errors on coefficients.}
 \tablenotetext{h}{Standard deviation.}
 \end{deluxetable}

\begin{deluxetable}{llcccccccc}
\tabletypesize{\scriptsize}
 \tablewidth{0pt}
 \tablecaption{Same as Table~\ref{tab_pwz}, but for triple band PWZ relations.\label{tab_pwz_3bands}}
 \tablehead{
 \multicolumn{1}{c}{Mode\tablenotemark{a}} &
 \multicolumn{1}{c}{Band\tablenotemark{b}} &
 \multicolumn{1}{c}{$\xi$\tablenotemark{c}} &
 \multicolumn{1}{c}{$a$\tablenotemark{d}} &
 \multicolumn{1}{c}{$b$\tablenotemark{e}} &
 \multicolumn{1}{c}{$c$\tablenotemark{f}} &
 \multicolumn{1}{c}{$\sigma_{a}$\tablenotemark{g}} &
 \multicolumn{1}{c}{$\sigma_{b}$\tablenotemark{g}} &
 \multicolumn{1}{c}{$\sigma_{c}$\tablenotemark{g}}&
 \multicolumn{1}{c}{$rms$\tablenotemark{h}}
}
 \startdata
FU   &V,B-I   & 1.34 &  -1.01 &  -2.54 &  0.08 & 0.01 & 0.03 & 0.01 & 0.05\\
FO   &V,B-I   & 1.34 &  -1.48 &  -2.74 &  0.08 & 0.03 & 0.07 & 0.01 & 0.03\\
Glob &V,B-I   & 1.34 &  -0.99 &  -2.42 &  0.08 & 0.01 & 0.02 & 0.01 & 0.05\\
FU   &V,B-R   & 2.02 &  -1.17 &  -2.72 &  0.02 & 0.01 & 0.04 & 0.01 & 0.07\\
FO   &V,B-R   & 2.02 &  -1.58 &  -2.78 &  0.04 & 0.04 & 0.08 & 0.01 & 0.04\\
Glob &V,B-R   & 2.02 &  -1.15 &  -2.58 &  0.03 & 0.01 & 0.03 & 0.01 & 0.07\\
FU   &J,B-R   & 0.57 &  -0.99 &  -2.44 &  0.13 & 0.01 & 0.02 & 0.01 & 0.03\\
FO   &J,B-R   & 0.57 &  -1.51 &  -2.75 &  0.12 & 0.03 & 0.05 & 0.01 & 0.02\\
\enddata
 \tablenotetext{a}{Pulsation mode.}
 \tablenotetext{b}{Photometric bands.}
 \tablenotetext{c}{Coefficient of the color term.}
 \tablenotetext{d}{Zero-point of the  relation.}
 \tablenotetext{e}{Coefficient of the logarithmic period.}
 \tablenotetext{f}{Coefficient of the metallicity term.}
 \tablenotetext{g}{Errors on coefficients.}
 \tablenotetext{h}{Standard deviation.}
 \end{deluxetable}

\begin{deluxetable}{llcccccccc}
\tabletypesize{\scriptsize}
 \tablewidth{0pt}
 \tablecaption{Coefficients of the predicted metal-independent, dual and triple band, optical PWZ relations for FU, FO and FU+FO pulsators.
\label{PWZindep}}
 \tablehead{
 \multicolumn{1}{c}{Mode\tablenotemark{a}} &
 \multicolumn{1}{c}{Band\tablenotemark{b}} &
 \multicolumn{1}{c}{$\xi$\tablenotemark{c}} &
 \multicolumn{1}{c}{$a$\tablenotemark{d}} &
 \multicolumn{1}{c}{$b$\tablenotemark{e}} &
 \multicolumn{1}{c}{$\sigma_{a}$\tablenotemark{f}} &
 \multicolumn{1}{c}{$\sigma_{b}$\tablenotemark{f}} &
 \multicolumn{1}{c}{$rms$\tablenotemark{g}}
}
 \startdata
FU  &V,B-R & 2.02 &  -1.20 &  -2.74 &   0.01 &   0.04 &   0.07\\
FO  &V,B-R & 2.02 &  -1.74 &  -3.00 &   0.04 &   0.10 &   0.05\\
Glob&V,B-R & 2.02 &  -1.20 &  -2.60 &   0.01 &   0.03 &   0.07\\
FU  &V,B-V & 3.06 &  -1.08 &  -2.66 &   0.01 &   0.04 &   0.07\\
FO  &V,B-V & 3.06 &  -1.56 &  -2.76 &   0.04 &   0.09 &   0.05\\
Glob&V,B-V & 3.06 &  -1.07 &  -2.49 &   0.01 &   0.04 &   0.08\\
FU  &R,B-R & 1.68 &  -1.24 &  -2.76 &   0.01 &   0.04 &   0.07\\
FO  &R,B-R & 1.68 &  -1.79 &  -3.00 &   0.04 &   0.10 &   0.05\\
Glob&R,B-R & 1.68 &  -1.24 &  -2.63 &   0.01 &   0.03 &   0.07\\

\enddata
 \tablenotetext{a}{Pulsation mode.}
 \tablenotetext{b}{Photometric bands.}
 \tablenotetext{c}{Coefficient of the color term.}
 \tablenotetext{d}{Zero-point of the relation.}
 \tablenotetext{e}{Coefficient of the logarithmic period.}
 \tablenotetext{f}{Errors on coefficients}
 \tablenotetext{g}{Standard deviation.}
 \end{deluxetable}